\pdfoutput=1
\documentclass[numberedappendix,apj]{emulateapj}
\usepackage{apjfonts}
\usepackage{xspace}
\usepackage{amsmath}
\def\MH2c{\, M_{\rm H_2,cell}}
\def\Msunpc2{\,\rm M_{\odot}\,pc^{-2}}

\def\HH{{{\rm H}_2}}

\newcommand{\Msol}{{\,M}_\odot}

\newcommand{\pc} {{\,\rm pc}} 
\newcommand{\K} {{\,\rm K}} 
\newcommand{\cc}{{\,\rm {cm^{-3}}}}

\renewcommand{\vec}[1]{{\mathbf #1}} 
\newcommand{\kmsec}{{\,\rm {km\,s^{-1}} }}

\def\Gyr{\,{\rm Gyr}}
\def\Myr{\,{\rm Myr}}
\newcommand{\kpc} {{\,\rm kpc}}

\newcommand{\vel}{\mbox{\boldmath$v$}}

\newcommand{\nab}{\mbox{\boldmath$\nabla$}}
\newcommand{\ra}[1]{\renewcommand{\arraystretch}{#1}}

\begin{document}
\shorttitle{On the interplay between star formation and feedback}
\slugcomment{{\em ApJ submitted}} 
\shortauthors{}

\title{On the interplay between star formation and feedback in galaxy formation simulations}

\author{
Oscar Agertz\altaffilmark{1,2,3} and Andrey V. Kravtsov\altaffilmark{2,3,4}}


\altaffiltext{1}{Department of Physics, University of Surrey, Guildford, GU2 7XH, United Kingdom {\tt o.agertz@surrey.ac.uk}}
\altaffiltext{2}{Department of Astronomy \& Astrophysics, The University of Chicago, Chicago, IL 60637 USA}
\altaffiltext{3}{Kavli Institute for Cosmological Physics, The University of Chicago, Chicago, IL 60637 USA}
\altaffiltext{4}{Enrico Fermi Institute, The University of Chicago, Chicago, IL 60637 USA}

\begin{abstract}
We investigate the star formation-feedback cycle  in cosmological galaxy formation simulations, focusing on progenitors of Milky Way (MW)-sized galaxies. We find that in order to reproduce key properties of the MW progenitors, such as semi-empirically derived star formation histories and the shape of rotation curves, our implementation of star formation and stellar feedback  requires 1) a combination of local early momentum feedback via radiation pressure and stellar winds and subsequent efficient supernovae feedback, and 2) efficacy of feedback that results in {\it self-regulation} of the global star formation rate on kiloparsec scales. We show that such feedback-driven self-regulation is achieved globally for a {\it local} star formation efficiency per free fall time of $\epsilon_{\rm ff}\approx 10\%$. Although this value is larger that the $\epsilon_{\rm ff}\sim 1\%$ value usually inferred from the Kennicutt-Schmidt (KS) relation, we show that it is consistent with direct observational estimates of  $\epsilon_{\rm ff}$ in molecular clouds. Moreover, we show that simulations with local efficiency of  $\epsilon_{\rm ff}\approx 10\%$ reproduce the global observed KS relation. Such simulations also reproduce the cosmic star formation history of the Milky Way sized galaxies and satisfy a number of other observational constraints. Conversely, we find that simulations that a priori assume an inefficient mode of star formation, instead of achieving it via stellar feedback regulation, fail to produce sufficiently vigorous outflows and do not reproduce observations. This illustrates the importance of understanding the complex interplay between star formation and feedback and the detailed processes that contribute to the feedback-regulated formation of galaxies. 
\end{abstract}

\setcounter{figure}{0}
\section{Introduction}
\label{sect:introduction}
The basic scenario of hierarchical galaxy formation \citep[][]{WhiteRees78,FallEfstathiou80} has been greatly elaborated and put on a firm footing within the Cold Dark Matter paradigm during the last three decades. Although the $\Lambda$ Cold Dark Matter ($\Lambda$CDM) model  has proven broadly successful in explaining and predicting a variety of observations, such as the Cosmic Microwave Background temperature anisotropies \citep[e.g][]{komatsu_etal11,hinshaw_etal13,Planckparam2013}, the evolution of cluster abundance \citep{vikhlinin_etal09b}, and large scale distribution of matter in the Universe \citep{conroy_etal06,Springel2006}, many aspects of the theory of galaxy formation are not yet fully understood \citep[see, e.g.,][for a recent review]{Silk2012}. 

One of the most pressing problems in galaxy formation modelling is understanding  why galaxies forming at the centers of dark matter halos are so inefficient in converting their baryons into stars. A number of different methods, such as dark matter halo abundance matching \citep{conroy_wechsler09,Guo2010}, satellite kinematics \citep{klypin_prada09,More2011}, and weak lensing \citep{Mandelbaum2006} \citep[see][for a comprehensive discussion]{kravtsov_etal14} point towards \emph{peak} stellar to dark matter mass fractions of $M_\star/M_{\rm h}\approx 3-5\,\%$ on average for $L_\star$ galaxies \citep[e.g.,][]{kravtsov_etal14}, far below the cosmological baryon fraction $\Omega_{\rm b}/\Omega_{\rm m}\approx 16\%$ \citep{Planckparam2013}. 

The low galactic baryon fractions are believed to be due to galactic winds driven by stellar feedback at the faint end of the stellar mass function \citep{DekelSilk86,Efstathiou00} and by the active galactic nuclei (AGN) and the bright end \citep{SilkRees1998,Benson2003}. Modeling these processes in fully cosmological hydrodynamical simulations has proven to be a daunting task due to the multi-scale nature of galaxy formation, where properties of the intergalactic distribution of baryons, on scales $\gtrsim 100\kpc$, are affected by star formation and feedback processes on scales of individual star clusters ($\lesssim 1\pc$). 

Although a formal spatial resolution of $\sim 10-100\pc$,  comparable to the scale of massive giant molecular clouds (GMCs), is not uncommon  in modern cosmological galaxy formation simulations \citep[e.g.][]{kravtsov03,GnedinKravtsov2010,Agertz09b,Hopkins2014}, the relevant star formation and feedback processes remain ``subgrid''. In particular,  substantial differences in resulting galaxies may arise when different implementations and paramterizations of these processes are used in simulations, 
even when the same initial conditions are used \citep{Governato10,Aquila}. 

Implementations of stellar feedback in galaxy formation simulations have been explored in many studies over the last two decades \citep[e.g.][]{Katz92,NavarroWhite93, Katz1996, ThackerCouchman2001, Stinson06, Governato07,Scannapieco08,Colin2010,Agertz2011,AvilaReese2011,Guedes2011,Aquila,Hopkins2011,Brook2012,Stinson2013,Agertz2013,Ceverino2013,Roskar2013,Booth2013,Christensen2014}. Nevertheless, we still do not have a full understanding of what processes matter most for suppressing star formation and driving galactic winds over the vast range of observed galaxy masses.

Recent studies \citep[][]{Leitner2012,Weinmann2012, Behroozi2013,Moster2013} have shown that not only is galaxy formation an inefficient process, but also that star formation in progenitors of most galaxies ($L\lesssim L_\star$) is significantly suppressed during the first 3 Gyr of cosmic evolution. \cite{vandokkum2013} recently reached a similar conclusion by matching cumulative co-moving number densities in the 3D-HST and CANDELS Treasury surveys, demonstrating that $\sim 90\%$ of the stellar mass in Milky Way mass galaxies formed after $z\sim2.5$. 

Much effort has gone into reproducing the $z=0$ $M_\star-M_{\rm h}$ relation over a large range of galaxy masses in simulations with efficient feedback \citep[e.g.,][]{Munshi2013}. At the same time, predicting its evolution, and hence reproducing the significant suppression of star formation necessary at $z\gtrsim2$ has proven more difficult. \cite{Brook2012} and \cite{Stinson2013} discussed the importance of ``early feedback''\footnote{Feedback that operates at times before the first SNe explosions, i.e. $t\lesssim 4\Myr$, for a coeval stellar population.} in their SPH simulations, here modeled by assuming that $10\%$ of the bolometric luminosity radiated by young stars get converted into thermal energy. This large energy injection resulted in star formation histories consistent with the data of \cite{Moster2013}. Similar results were found by \cite{Aumer2013} who considered a momentum based model of radiation pressure, 
although with the value of the infrared optical depth of $\tau_{\rm IR}\sim 25$, larger than in the models by \cite{Hopkins2011} and \cite{Agertz2013}. \cite{Hopkins2014}, \cite{Ceverino2013}, and  \cite{TrujilloGomez2013} also found that radiative feedback, both due to photoionizaiton and radiation pressure, could play an important role in low mass galaxies (here progenitors of galaxies with $M_{\rm vir}(z=0)\lesssim 10^{12}\Msol$) at high redshifts, even for more moderate values of photon trapping by dust. 

While suppression of star formation in simulations of galaxy formation via strong stellar feedback has been widely  explored in the recent literature, freedoms in the way in which star formation in the interstellar medium (ISM) is modeled has received less attention. Recent work by \citet[][see also \citealt{GnedinKravtsov2010,GnedinKravtsov11,Kuhlen2012,Christensen2014}]{Gnedin09} demonstrated how a star formation model based on the local abundance of $\HH$ could explain the observed steepening for $\Sigma_{\rm gas}<100\Msol\pc^{-2}$ in the Kennicutt-Schmidt (KS) relation for $z\approx 3-4$ Damped Lyman-$\alpha$ systems and Lyman Break Galaxies (LBGs). \cite{Governato10} found that a high threshold for star formation, in conjunction with higher resolution and strong feedback, can lead to more correlated feedback events and a more realistic halo baryon fraction. These results illustrate that it is paramount to explore how parameters of the star formation recipe and feedback implementation affect basic properties of galaxies forming in a given halo. 

In this paper we present results of a systematic study of such dependencies using high resolution, cosmological simulations of the Milky Way (MW) sized progenitors that include our new model for stellar feedback described in \cite{Agertz2013}. We specifically explore how the interplay between various modes of star formation and feedback models affect galactic characteristics at $z\gtrsim 1$. The paper is organized as follows: In \S\,\ref{sect:method} we outline our numerical method as well as star formation and feedback models. In \S \ref{sec:sfeff} we discuss empirical constraints on the efficiency of star formation in molecular clouds -- one of the most important parameters in our implementation of the star formation -- feedback cycle, and show that observations often indicate an efficiency in massive star forming clouds considerably larger than implied by the global normalization of the Kennicutt-Schmidt relation. \S\,\ref{sect:IC}
describes the initial conditions and the simulation suite. In \S\ref{sect:results} we present our suite of cosmological simulations and demonstrate how two different models of star formation and feedback can match several observational properties of galaxies, including the star formation history, the total stellar mass expected from abundance matching, average gas metallicity and the rotational velocity. In \S\,\ref{sect:degeneracy} we discuss how the degeneracy between the two parameterizations can be broken, and show how only the simulation with efficient stellar feedback together with a high local efficiency of star formation can reproduce all the observed properties of galaxies. Finally, we discuss our results and conclusions in \S\,\ref{sect:discussion} and \ref{sect:conclusions}.

\section{Numerical code}
\label{sect:method}

We carry out cosmological hydro+$N$-body simulations using the Adaptive Mesh Refinement (AMR) code {\tt RAMSES} \citep{teyssier02}. The fluid dynamics of baryons is calculated using a second-order unsplit Godunov method, while the collisionless dynamics of stellar and dark matter particles is evolved using the particle-mesh technique with gravitational accelerations computed from the gravitational potential on the mesh. The gravitational potential is calculated by solving the Poisson equation using the multi-grid method \citep{brandt77,GuilletTeyssier2011} for all refinement levels. The potential is used to compute accelerations for both the particles and the baryon fluid. The equation of state of the fluid is that of an ideal mono-atomic gas with an adiabatic index $\gamma=5/3$. 

The code achieves high resolution in high density regions using adaptive mesh refinement, where the refinement strategy is based on a quasi-Lagrangian approach in which the number of collisionless particles per cell is kept approximately constant. This allows the local force softening to closely match the local mean interparticle separation, which suppress discreteness effects \citep[e.g.,][]{knebe_etal00,Romeo08}. An analogous refinement criterion is also used for the gas. 

\subsection{Star formation}
\label{sect:SF}
We model the local star formation rate using the following equation:
\begin{equation}
\label{eq:schmidtH2}
\dot{\rho}_{\star}=f_{\rm H_2}\frac{ \rho_{\rm g}}{t_{\rm SF}}, 
\end{equation}
where $f_{\rm H_2}$ is the local mass fraction of molecular hydrogen (H$_2$), $\rho_{\rm g}$ is the gas density in a cell, and $t_{\rm SF}$ is the star formation time scale of {\it molecular} gas. In \S\,\ref{sect:h2model} we describe the model we use to calculate $f_{\rm H_2}$. The time scale $t_{\rm SF}$ is defined by the efficiency of star formation, which, as we show below, is one of the key parameters controlling basic properties of galaxies forming in a given halo and efficacy of stellar feedback. Given its importance, we will discuss the empirical constraints and our choices for the value of this parameter in our simulations in \S~\ref{sec:sfeff} below. 

To ensure that the number of star particles formed during the course of a simulation is tractable, we sample Equation~\ref{eq:schmidtH2} stochastically at every fine simulation time step $\Delta t$ \citep[see section 2.3 in][for details]{Agertz2013}. We also adopt a temperature threshold by only allowing star formation to occur in cells of  $T<10^4\K$, although we find that this threshold has no actual impact on the resulting star formation rates.

By adopting the kind of star formation relation in Equation\,\ref{eq:schmidtH2}, we avoid imposing a fixed, and perhaps arbitrary, star formation density threshold, as is common in the galaxy formation community. We explore the difference between the constant density threshold approach and the molecular hydrogen based star formation model further Appendix\,\ref{appendix:A}. As described in \S\,\ref{sect:introduction}, relating star formation to the molecular gas is well motivated empirically, as galactic star formation rate surface densities correlate well with the surface density of molecular gas independent of metallicity, and poorly or not at all with the surface density of atomic gas measured on kpc scales \citep{bigiel2008,Krumholz09,Gnedin09}. 

\subsection{Molecular hydrogen model}
\label{sect:h2model}
To capture the physics of molecular gas in our simulations, which is a key ingredient in our star formation model (see \S\,\ref{sect:SF}), we adopt the KMT09 model, which we briefly discuss in this section. Molecular hydrogen forms readily when dust grains are present, but the abundance is very sensitive to the destruction by UV radiation. \cite{kmt08}, \cite{kmt09}, and \cite{McKeeKrumholz2010} developed a model for the abundance of $\HH$ based on radiative transfer calculations of idealized spherical giant atomic--molecular complexes subject to a uniform and isotropic Lyman-Werner (LW) radiation field. When the  $\HH$ abundance is calculated assuming formation-dissociation balance, the solution can conveniently be expressed as
\begin{equation}
f_\HH \simeq 1 - \frac{3}{4}\frac{s}{1 + 0.25 s}, \label{eq:KMT09model_begin}
\end{equation}
\begin{equation}
s = \frac{\ln(1 + 0.6\chi + 0.01 \chi^2)}{0.6\,\tau_c},
\end{equation}
\begin{equation}
\label{eq:chi}
\chi = 71 \left( \frac{\sigma_{d,-21}}{\mathcal{R}_{-16.5}} \right) \frac{G'_0}{n_{\rm H}},
\end{equation}
where $\tau_c$ is the dust optical depth of the cloud, $\sigma_{d,-21}$ is the dust cross-section per hydrogen nucleus to radiation at 1000 \AA\ normalized to $10^{-21}$ cm$^{-2}$ and $n_{\rm H}$ is the volume density of hydrogen nuclei in units of cm$^{-3}$. The coefficient $\mathcal{R}_{-16.5}$ is the rate for $\HH$ formation on dust grains, normalized to the Milky Way value of $10^{-16.5}$ cm$^3$ s$^{-1}$ \citep[see][]{Wolfire2008} and  $G'_0$ is the ambient UV radiation field intensity, normalized to the \cite{Draine1978} value for the Milky Way.  As both $\sigma_d$ and $\mathcal{R}$ are linearly proportional to the dust abundance, and hence gas metallicity, their ratio in $\chi$ becomes independent of metallicity.

The equations above can be simplified further by assuming pressure equilibrium between the cold and warm neutral medium (CNM and WNM respectively). \cite{kmt09} demonstrated that the assumption of pressure balance between the two gas phases causes the minimum CNM density to be linearly proportional to the UV radiation field:
\begin{equation}
n_{\rm min}\approx\frac{31}{1+3.1Z_{\rm g}^{0.365}}G'_0,
\end{equation}
where $Z_{\rm g}$ is the gas metallicity in units of solar metallicity, $Z_\odot=0.020$. By allowing for the CNM density to be larger than the minimum density by a factor $\phi_{\rm CNM}$, i.e. $n_{\rm H}=\phi_{\rm CNM} n_{\rm min}$, Equation\,\ref{eq:chi} becomes
\begin{equation}
\label{eq:chi2}
\chi = 2.3 \left( \frac{\sigma_{d,-21}}{\mathcal{R}_{-16.5}} \right) \frac{1 + 3.1 \, Z_{\rm g}^{0.365}}{\phi_{\rm CNM}}.
\end{equation}
As seen in the equation above, the molecular hydrogen mass fraction becomes independent of the local LW intensity. \cite{krumholzgnedin2011} found that the two phase approximation predicts the H$_2$ abundance accurately compared to full non-equilibrium radiative transfer calculations for $Z_{\rm g}\gtrsim 10^{-2}Z_\odot$.

In the remainder of the paper we refer to the above model, including the two-phase CNM-WNM equilibrium assumption, as the KMT09 model. The KMT09 model was adopted in fully cosmological simulations of galaxy formation by \cite{Kuhlen2012} and \cite{Kuhlen2013} \citep[see also][]{Tomassetti2015}, who demonstrated how the model led to a strong suppression of star formation in low-mass halos ($M_{\rm h}\lesssim10^{10}\Msol$) at $z>4$, in agreement with galaxy formation simulations of \citet{GnedinKravtsov2010} which used full non-equilibrium calculations of H$_2$ abundance.

\subsection{Feedback}
\label{sect:FB}
The stellar feedback model adopted in our simulations is described in detail in \cite{Agertz2013}. Briefly, each formed stellar particle is treated as a single-age stellar population with a \cite{chabrier03} initial mass function (IMF).  Several processes are contributing to stellar feedback, as stars inject energy, momentum, mass and heavy elements over time via SNII and SNIa explosions, stellar winds and radiation pressure into the surrounding gas. Hence, at every simulations time step, and for every stellar particle, we account for the following energy, momentum, mass loss and metal injection rates:
\begin{eqnarray}
\mbox{\emph{Energy:}}&\quad \dot{E}_{\rm tot} &=  \dot{E}_{\rm SNII}+\dot{E}_{\rm SNIa}+\dot{E}_{\rm wind} \nonumber \\
\mbox{\emph{Momentum:}}&\quad \dot{p}_{\rm tot} &=  \dot{p}_{\rm SNII}+\dot{p}_{\rm wind}+\dot{p}_{\rm rad}\label{eq:FBarray} \\
\mbox{\emph{Mass loss:}}&\quad \dot{m}_{\rm tot} &=  \dot{m}_{\rm SNII}+\dot{m}_{\rm SNIa}+\dot{m}_{\rm wind}+\dot{m}_{\rm loss} \nonumber\\ 
\mbox{\emph{Metals:}}&\quad \dot{m}_{\rm Z,tot} &=  \dot{m}_{\rm Z,SNII}+\dot{m}_{\rm Z,SNIa}+\dot{m}_{\rm Z,wind}+\dot{m}_{\rm Z,loss},\nonumber
\end{eqnarray}
Each term in the above equations depends on the stellar age, mass and gas/stellar metallicity, all accounted for and described in \cite{Agertz2013}. Feedback is thus not done instantaneously, but continuously at the appropriate times when the various feedback process are known to operate, taking into account the lifetime of stars of different masses in a stellar population. To track the lifetimes of stars within the population we adopt the approximation of the metallicity dependent age-mass relation of \cite{Raiteri1996}, obtained as a fit to the results of the Padova stellar evolution models \citep{alongi_etal93,bressan_etal93}. 

The effect of radiation pressure is modeled as a direct injection of momentum to the cells surrounding newly formed star particles. Here the momentum injection rate from radiation can be written as
\begin{equation}
\label{eq:radpressure}
\dot{p}_{\rm rad}=(\eta_1+\eta_2\tau_{\rm IR})\frac{L(t)}{c},
\end{equation}
where $\tau_{\rm IR}$ is the infrared optical depth and $L(t)$ is the luminosity of the stellar population, here taken from the stellar evolution code {\tt STARBURST99} \citep{Leitherer1999}. The first term describes the direct radiation absorption/scattering, and given the large dust and HI opacities in UV present in dense star forming regions, $\eta_1\approx 1$. The second term describes momentum transferred by infrared photons re-radiated by dust particles, and scattered multiple times by dust grains before they escape, where $\eta_2$ is added to scale the fiducial value of $\tau_{\rm IR}$. Following \cite{Agertz2013}, we adopt $\eta_2=2$. As cosmological simulations cannot resolve the density structure around young massive star clusters on sub-parsec scales, to estimate $\tau_{\rm IR}$ we use the empirically-motivated subgrid model described in \cite{Agertz2013}. 

The momentum due to stellar winds, radiation pressure, and SN blastwaves is added to the 26 nearest cells surrounding parent cell of the stellar particle. The thermal energy due to SNe and shocked stellar winds is injected directly into the parent cell.

In most of our simulations we explore the concept of retaining some fraction of the thermal feedback energy in a \emph{separate} gas energy variable over longer times than expected purely form the local gas cooling time scale. This approach was discussed by \cite{Agertz2013}, and previously by \cite{Teyssier2013}, and can be viewed as accounting for the effective pressure from a multiphase medium, where local unresolved pockets of hot gas exert work on the surrounding cold phase, or a placeholder for other sources of energy, such as turbulence and cosmic rays \citep{Booth2013}. 

As described in section 3.2 of \cite{Agertz2013}, at each time step $\Delta t$ we inject a fraction $f_{\rm fb}$ of the calculated feedback energy into a separate energy variable, $E_{\rm fb}$, and the remaining energy fraction, $1-f_{\rm fb}$, is released as thermal energy into the main energy variable. In this work we adopt $f_{\rm fb}=0.5$. $E_{\rm fb}$ has units of energy per unit volume and evolves according to the following equation\footnote{note that this differs from the original implementation in \cite{Agertz2013} who neglected the $-P_{\rm fb}\nab\cdot\vel$ term. In \cite{Agertz2013}, the adiabatic cooling via $pdV$ work done by the \emph{total} pressure was affecting only the thermal energy component.}:
\begin{equation}
\label{eq:fbeq}
\frac{\partial}{\partial t}(E_{\rm fb})+\nab\cdot(E_{\rm fb}\vel_{\rm gas} )=-P_{\rm fb}\nab\cdot\vel_{\rm gas}-\frac{E_{\rm fb}}{t_{\rm dis}}.
\end{equation}
Note that $E_{\rm fb}$ refers to the variable followed by the above equation, not to be confused with $E_{\rm SNII}$ that denotes energy released by type II SNe. In the momentum equation, the thermal pressure $P_{\rm therm}$ is replaced by the total pressure $P_{\rm tot}=P_{\rm therm}+P_{\rm fb}$, where $P_{\rm fb}=(\gamma-1)E_{\rm fb}$. To achieve numerical stability, the Courant-Friedrichs-Lewy (CFL) condition is also updated to account for the sound speed related to the new total pressure when computing the simulations time step $\Delta t$. When stellar feedback is vigorous, we find that $\Delta t$ can be as low as $\sim 500-1000$ years.

As seen from Equation\,\ref{eq:fbeq}, the feedback energy is thus continuously dissipated over a time-scale $t_{\rm dis}$, i.e. $E_{\rm fb}^{t+\Delta t}=E_{\rm fb}^t\exp{(-\Delta t/t_{\rm dis}})$. We make the assumption that the dissipation timescale is comparable to the decay time of supersonic turbulence, which is of order of the flow crossing time \citep{Ostriker2001}. In all of the simulations presented in this paper, we adopt a fixed $t_{\rm dis}=10\,\Myr$, typical for a few crossing times in massive GMCs ($l\sim 10-100$~pc), or the vertical crossing time in cold galactic disks, with characteristic velocity dispersions $\sigma_{\rm HI}\sim 10\kmsec$. 

Heavy elements (metals) injected by supernovae and winds are advected as a passive scalar and are incorporated self-consistently in the cooling and heating routine. We adopt the tabulated cooling functions of \cite{sutherlanddopita93} for cooling at temperatures $10^4-10^{8.5}\,$K, and extend cooling down to $T=300\K$ using rates from \cite{rosenbregman95}. Heating from the UV background (UVB) radiation is accounted for by using the UVB model of \cite{haardtmadau96}, assuming a reionization redshift of $z=8.5$. We follow \cite{Agertz09b} and adopt an initial metallicity of $Z=10^{-3}Z_\odot$ in the high-resolution region (see \S\ref{sect:IC}) in order to account for enrichment from unresolved Pop III stars \citep[e.g.][]{wise_etal12}; their effect needs to be accounted for as it allows for the first molecular hydrogen to be synthesized in high-$z$ galaxy progenitors, hence initiating star formation. Note that the dependence of $f_{\rm H_2}$ on metallicity at $Z/Z_{\odot}\lesssim 10^{-2}$ is not known and is subject to effects such as Lyman-Werner band line overlap \citep{gnedin_draine14}. Thus, our assumption about the metallicity floor is within the uncertainties of the Population III SNe and $f_{\rm H_2}$ modelling.

\begin{figure}[t]
\begin{center}
\includegraphics[width=0.5\textwidth]{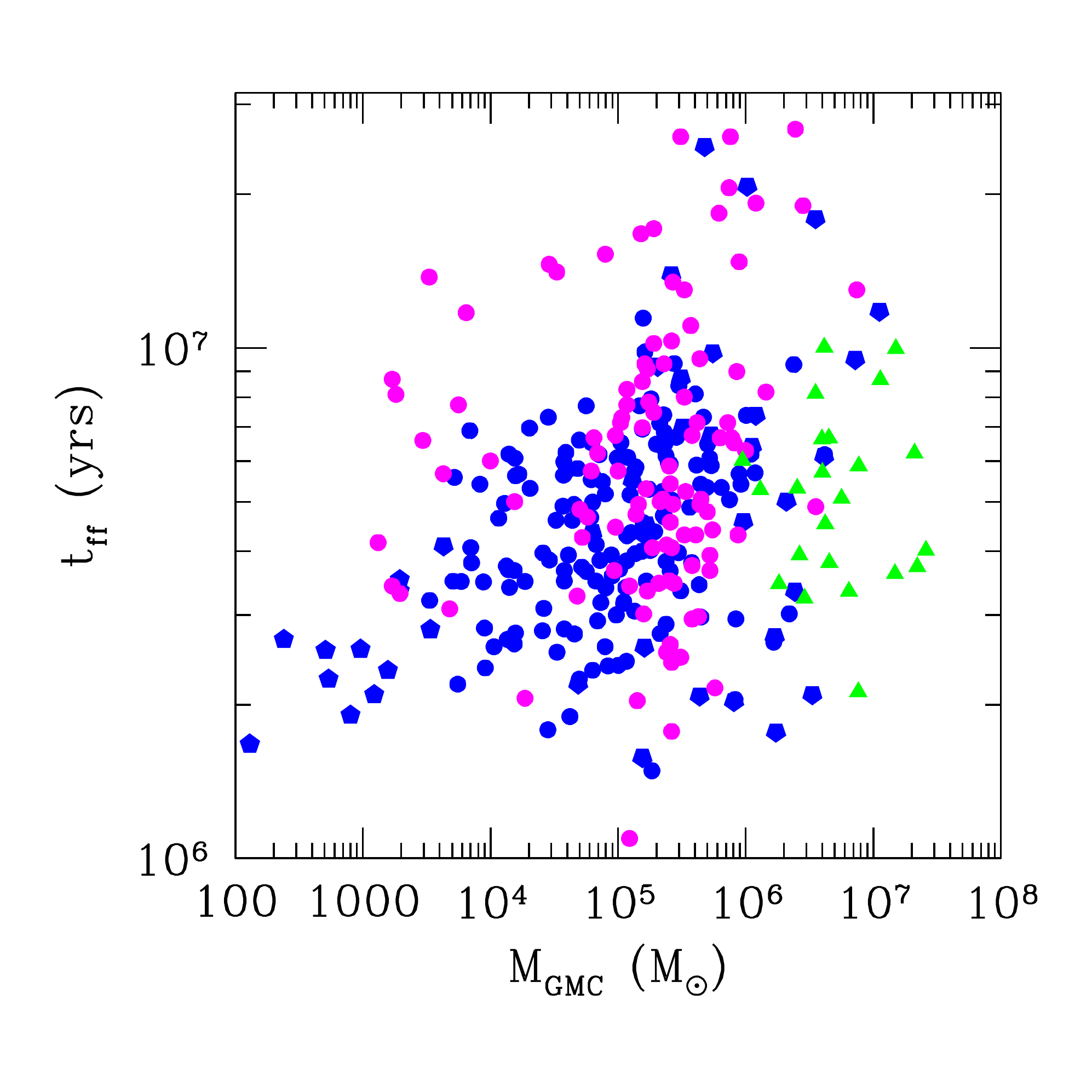}
\caption{Mean free-fall times of giant molecular clouds in different samples and galaxies. {\it Blue circles} show GMCs from the sample of \protect\cite{heyer_etal09}, {\it blue pentagons} are GMCs presented in Table 1 of \protect\cite{Murray2011b}, {\it magenta circles} show GMCs in several nearby galaxies including the Milky Way from the sample of \protect\cite{bolatto_etal08}, and {\it green triangles} show GMCs in the dense molecular inner region of M64 in the sample of \protect\cite{rosolowsky_blitz05}.}
\label{fig:tff}
\end{center}
\end{figure}
\begin{figure}[t]
\begin{center}
\includegraphics[width=0.5\textwidth]{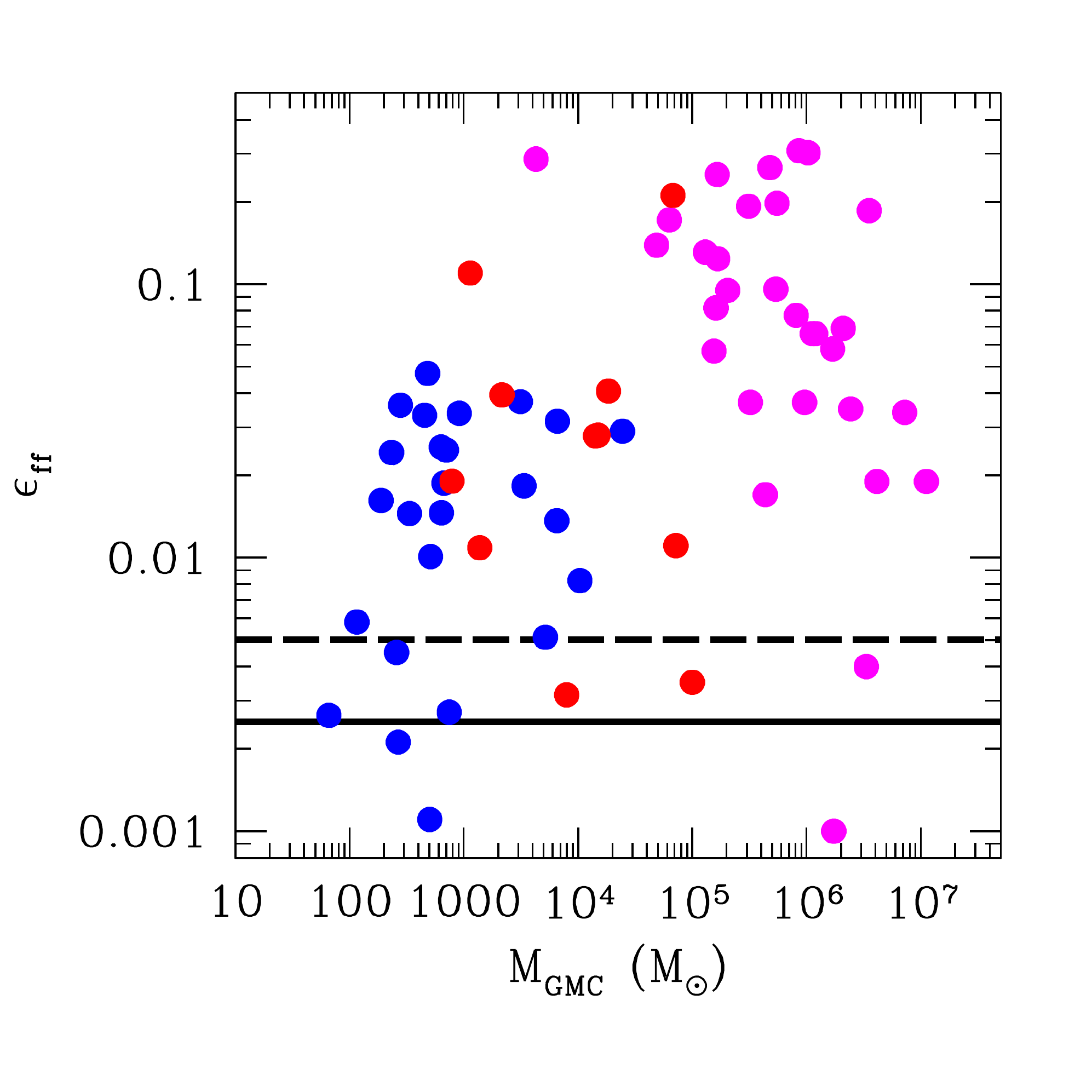}
\caption{The {\it local} efficiency per free-fall time for individual giant molecular clouds in the samples of \protect\citet[][blue points]{evans_etal14}, \protect\citet[][red points]{lada_etal10} and \protect\citet[][magenta points]{Murray2011}. The solid line shows the value of $\epsilon_{\rm ff}$ implied by the median global molecular gas consumption scale of $\tau_{\rm H_2}=2$ Gyrs inferred from the Kennicutt-Schmidt relation \citep{bigiel2008}, while the dashed line shows a similar estimate for $\tau_{\rm H_2}\approx 1$ Gyr, inferred specifically for the Milky Way. }
\label{fig:eff}
\end{center}
\end{figure}

\section{Efficiency of star formation}
\label{sec:sfeff}
The star formation time scale of {\it molecular} gas, $t_{\rm SF}$, in our adopted  star formation relation (Equation \ref{eq:schmidtH2}) is related to the {\it local} efficiency of star formation in a computational cell of a given density. Following \cite{krumholztan07}, we can write this time scale as $t_{\rm SF}=t_{\rm ff, SF}/\epsilon_{\rm ff, SF}$, where $t_{\rm ff, SF}=\sqrt{3\pi/32G\rho_{\rm g}}$ is the local free-fall time of the star forming gas and $\epsilon_{\rm ff, SF}$ is the {\it local} star formation efficiency per free-fall time. As we show below, basic properties of galaxies forming in a given halo, and the degree to which these properties are affected by stellar feedback, depend sensitively on the value of $t_{\rm SF}$ or $\epsilon_{\rm ff, SF}$. It is therefore important to discuss the motivation behind particular values of this parameter that we adopt in our simulations. 

Star formation overall, and the efficiency with which a given molecular region converts its gas mass into stars, are not yet fully understood theoretically. Nevertheless, useful empirical constraints do exist, and a plethora of theoretical models predicting the star formation efficiency have been developed over the last decade \citep[][]{Padoanreview2013}. 

On global, kiloparsec scales observational measurements show that the gas consumption time scale of molecular gas is $t_{\rm H_2,\,gal}\approx 2$ Gyrs \citep{bigiel2008}. If $\epsilon_{\rm ff}$ had a universal value in all of the molecular gas, and molecular gas had a common characteristic free-fall time, we would expect a direct relation between the global molecular gas consumption time scale and the local gas consumption time in star forming clouds, i.e. $t_{\rm SF}\approx t_{\rm H_2,\,gal}$ and thus $\epsilon_{\rm ff, SF}\approx t_{\rm ff,SF}/t_{\rm H_2,\,gal}$. 

Figure~\ref{fig:tff} shows estimates of the free-fall time for individual star forming GMCs from samples in both the Milky Way and other galaxies collected from the literature, as described in the caption. For each GMC, $t_{\rm ff, GMC}$ was computed as $t_{\rm ff, GMC}=\sqrt{3\pi/32G\rho_{\rm GMC}}$, where $\rho_{\rm GMC}=3M_{\rm GMC}/(4\pi R^3)$ is the mean density of the GMC computed using its mass and density reported in the corresponding sample. 
The figure shows that the gas consumption time scale in star forming clouds, $t_{\rm ff, SF}\sim t_{\rm ff,\, GMC}$, does not depend on GMC mass. The median value is $t\sim 5\times 10^6$ yrs, but the scatter in individual values is quite large, with values spanning the range from $\approx 10^6$ to $3\times 10^7$ years. If all molecular gas was in such GMCs, these values would imply values of $\epsilon_{\rm ff, GMC}\approx t_{\rm ff,GMC}/t_{\rm H_2,\, gal}$ from $0.0005$ to $0.015$, with a median of $\approx 0.0025$. The long global consumption time scale of the molecular gas,  $t_{\rm H_2,\,gal}\approx 2$ Gyrs,  thus, is often taken to indicate low star formation efficiencies in star forming clouds. 

However, the values of $\epsilon_{\rm ff}$ estimated in this way are significantly lower than estimates for individual star forming regions, which have typical values in the range of $\epsilon_{\rm ff,\, GMC}\sim 0.01-0.1$ \citep{lada_etal10,Murray2011b,evans_etal14}. This is illustrated in Figure~\ref{fig:eff}, which shows values of $\epsilon_{\rm ff, GMC}$ estimated for  samples of individual GMCs in the Milky Way by \cite{Murray2011} and \cite{evans_etal14} as a function of GMC mass. We have also included GMCs analyzed by \cite{lada_etal10} for which we have estimated $\epsilon_{\rm ff, GMC}$ assuming a fixed free-fall time of $5\times 10^6$ yrs corresponding to the median free-fall time of GMCs in Figure~\ref{fig:tff}.

Note that numerical simulations of turbulent bound star forming clouds \citep[e.g., see][for a recent review]{Padoanreview2013} also generally predict values of $\epsilon_{\rm ff}$ much larger than 0.0025. Such theoretical predictions may thus explain the high values of $\epsilon_{\rm ff}$ in a fraction of the most bound, massive GMCs. 
 
The sizeable difference in values of $\epsilon_{\rm ff}$ derived from the global molecular gas consumption time scale and direct estimates of the {\it local} efficiency, $\epsilon_{\rm ff, GMC}$, in star forming clouds indicates that $t_{\rm SF}\ne t_{\rm H_2,\,gal}$, and thus a significant fraction of molecular gas is not star forming, or forms stars with an extremely low efficiency. This is corroborated by measurements of the molecular gas consumption time scale distribution of molecular gas in patches of 12 pc radius in the SMC by \cite{bolatto_etal11}. The typical consumption time scale for molecular gas is indeed several billion years, and a high star formation efficiency is reached only in a small fraction of molecular patches. 

These considerations indicate that the local value of $\epsilon_{\rm ff, SF}$ in star forming regions on small scales does not have to correspond to the global value implied by the molecular gas consumption time scale on kiloparsec scales and can be significantly larger. In our study we therefore consider a range of values of $\epsilon_{\rm ff}$ from $0.01$ to $0.1$, consistent with the empirical estimates of this efficiency in GMCs shown in Figure~\ref{fig:eff}. 

\begin{table*}[t]
\begin{center}
\caption{List of cosmological zoom-in simulations. All simulations reach a minimum cell size of $\Delta x=75$~pc.}
\label{table:simsummary1}
\begin{minipage}{140mm}
\center
\ra{1.3}
\begin{tabular}{ll}
\hline
\hline \\[-5pt]
Simulation & Description \\ 
\\[-5pt]
\hline\\[-8pt]
{\it KMT09 models} & \\
\\[-8pt]
\hline
{\tt NoFB\_e001} & No feedback (only metal enrichment), $\epsilon_{\rm ff}=1\%$\\
{\tt NoFB\_e010} & No feedback (only metal enrichment), $\epsilon_{\rm ff}=10\%$\\
{\tt ALL\_e010} & All feedback processes, $\epsilon_{\rm ff}=10\%$\\
\hline \\[-8pt]
{\it KMT09 models, feedback energy variable $E_{\rm fb}$, $f_{\rm fb}=0.5$, $t_{\rm dis}=10\Myr$ }& \\
\\[-8pt]
\hline
{\tt ALL\_Efb\_e010} & All feedback processes, $\epsilon_{\rm ff}=10\%$\\
{\tt NoPrad\_Efb\_e010} & All feedback processes but radiation pressure, $\epsilon_{\rm ff}=10\%$\\
{\tt ALL\_Efb\_e001} & All feedback processes, $\epsilon_{\rm ff}=1\%$\\
{\tt ALL\_Efb\_e001\_5ESN} & All feedback processes, $E_{\rm SNII}=5\times 10^{51}\,{\rm erg}$, $\epsilon_{\rm ff}=1\%$
\\
\hline \\[-8pt]
{\it Fixed threshold for star formation ($n_\star=25\cc$), $E_{\rm fb}$, $f_{\rm fb}=0.5$, $t_{\rm dis}=10\Myr$ }& \\
\\[-8pt]
\hline
{\tt ALL\_Efb\_e010\_n25} (see appendix A) & All feedback processes, $\epsilon_{\rm ff}=10\%$\\
\\[-8pt]
\hline

\end{tabular}
\end{minipage}
\end{center}
\end{table*}

\section{Initial conditions and simulation suite} 
\label{sect:IC}

The initial conditions used in this work are identical to those presented in \cite{Agertz2011}. In summary, we adopt a \emph{WMAP5} \citep{Komatsu2009} compatible $\Lambda$CDM cosmology with $\Omega_{\Lambda}=0.73$, $\Omega_{\rm m}=0.27$, $\Omega_{\rm b}=0.045$, $\sigma_8=0.8$ and $H_0=70\,{\rm km\,s}^{-1}\,{\rm Mpc}^{-1}$.  A pure dark matter simulation was performed using a simulation cube of size $L_{\rm box}=179\,{\rm Mpc}$. At $z=0$, a halo of mass $M_{\rm 200c}\approx 9.7\times10^{11}\,\Msol$ was selected for re-simulation at high resolution, and traced back to the initial redshift of $z=133$. Here $M_{\rm 200c}$ is  defined as the mass enclosed within a sphere with mean density 200 times the critical density at the redshift of analysis. The corresponding radius is $r_{\rm 200c}=205\,\kpc$. The mass within the radius enclosing overdensity  of 200 times the mean density is $M_{\rm 200m}=1.25\times10^{12}\,\Msol$ and $r_{\rm 200m}=340\kpc$. When baryons are included in the simulations, the final \emph{total} halo mass remains approximately the same.

The selected halo does not experience any major merger after $z=1$, favouring the formation of an extended late-type galaxy. A nested hierarchy of initial conditions for the dark matter and baryons was generated using the {\small GRAFIC++}\footnote{{\tt http://grafic.sourceforge.net/}} code, where we allow for the high resolution particles to extend to three virial radii from the centre of the halo at $z=0$. This avoids mixing of dark matter particles with different masses in the inner parts of the domain. The dark matter particle mass in the high resolution region is $m_{\rm DM}=3.2\times 10^5\,\Msol$ and the adaptive mesh is allowed to refine if a cell contains more than eight dark matter particles, and a similar criterion is employed for the baryonic component. At the maximum level of refinement, the simulations reach a physical resolution of $\Delta x\approx75\,\pc$.

\begin{figure*}[t]
\begin{center}
\includegraphics[width=1.\textwidth]{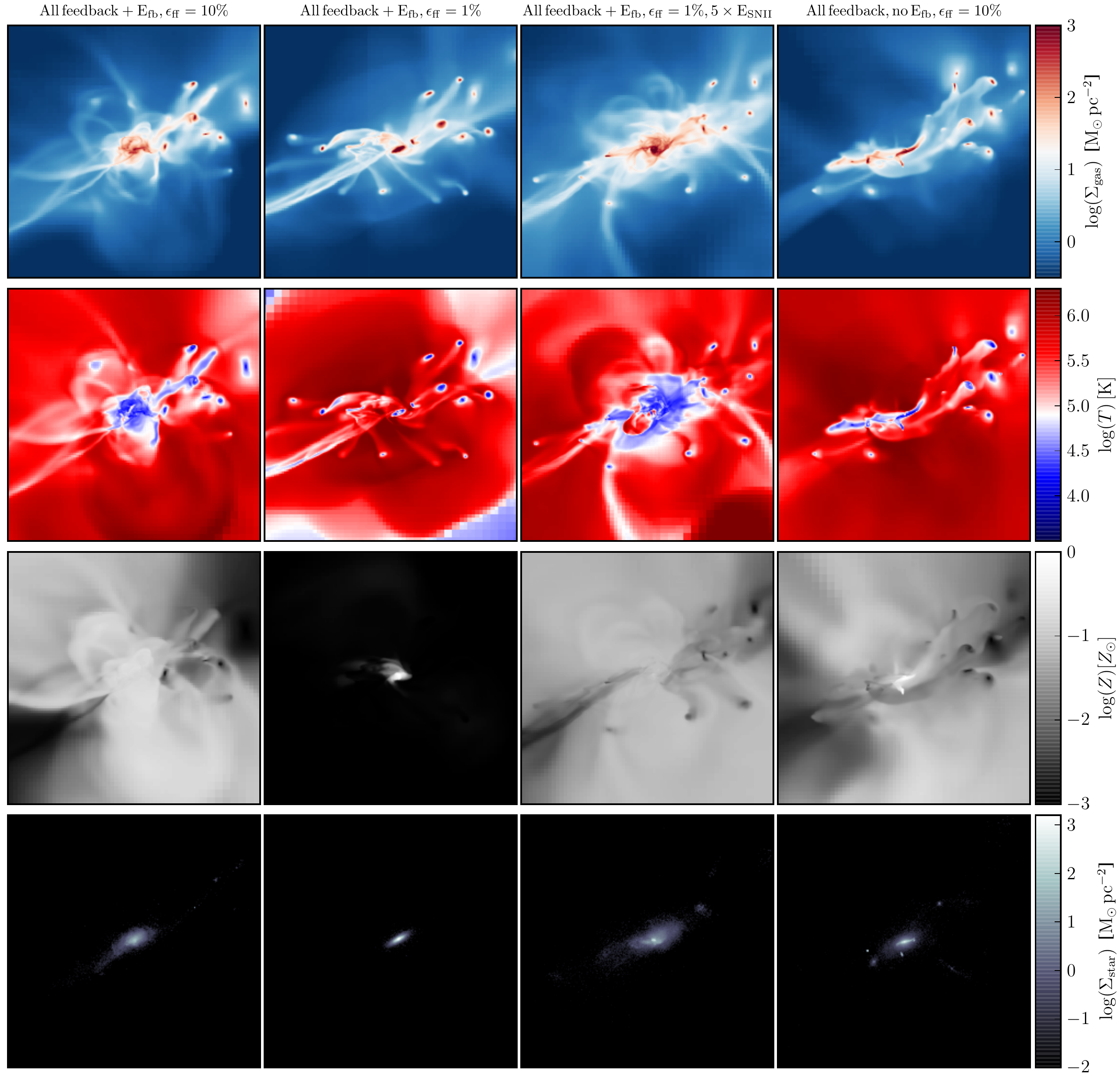}
\caption{Maps of, from top to bottom, gas surface density, mass weighted temperature, gas metallicity and stellar surface density for, from left to right, {\tt ALL\_Efb\_e010}, {\tt ALL\_Efb\_e001}, {\tt ALL\_Efb\_e001\_5ESN} and {\tt ALL\_e010} at $z=3$. The maps show regions 90 kpc on each side. All simulations, apart from {\tt ALL\_Efb\_e001}, show clear signatures of outflows. As discussed in the text, the low input free-fall time efficiency of star formation ($\epsilon_{\rm ff}=1\%$) does not allow for local feedback to be vigorous enough to generate galactic winds. }
\label{fig:map1}
\end{center}
\end{figure*}

\subsection{Simulation suite}
The main focus of this work is to investigate the interplay between star formation and feedback. To this end, we carry out a suite of simulations targeting a number of different regimes; 1) no stellar feedback from young stars, 2) all sources of stellar feedback are operating (as discussed in \S\,\ref{sect:FB}), 3) the impact of neglecting radiative feedback and 4) the impact of making SN energy  feedback less efficient by not tracking it as a separate fluid variable. 

For the first two regimes we also study the impact of varying the efficiency of star formation per-free-fall time (see \S \ref{sect:SF}) using simulations with $\epsilon_{\rm ff}=1\%$ and 10\%. The lower efficiency is closer to the value derived from the gas consumption time scale in kpc-sized patches of the ISM (see \S~\ref{sec:sfeff} above). However, the relevant values of $\epsilon_{\rm ff}$ for  GMCs as a function of environment is not fully understood, as we discussed above in \S~\ref{sec:sfeff} (see Figure~\ref{fig:eff}).  Simulating galaxy formation using larger efficiency of $\epsilon_{\rm ff}=10\%$ is thus motivated by GMC observations and allows us to study the ability for stellar feedback to regulate the \emph{measured} efficiency to globally observed values. 

Our fiducial simulations include all feedback process discussed above, including the second energy variable and star formation efficiency of $\epsilon_{\rm ff}=10\%$.  For the case of $\epsilon_{\rm ff}=1\%$, we also investigate the effect of increasing the available feedback energy from SNII events, going from the fiducial  $E_{\rm SNII}=10^{51}\,{\rm erg}$ to $5\times E_{\rm SNII}$. Such an increase could correspond to a somewhat more top heavy IMF. 

In Appendix\,\ref{appendix:A} we compare the results from our simulations with H$_2$ based star formation with similar simulation in which star formation is assumed to proceed at densities above a fixed density threshold. For the latter simulation we adopt a density threshold of $n_\star=25\cc$, which roughly corresponds to the physical density at which $f_{\HH}\sim 50\%$ at $Z_{\rm g}=Z_\odot$. We note that this threshold value is larger than the value adopted in the Eris simulation \citep{Guedes2011}, where $n_\star=5\cc$ was used, although close to the  $n_\star=20\cc$ adopted for the followup Eris2 simulation.

The entire simulation suite, and the associated star formation and feedback parameters, are summarized in Table \ref{table:simsummary1}.

\begin{figure*}[t]
\begin{center}
\includegraphics[scale=0.75]{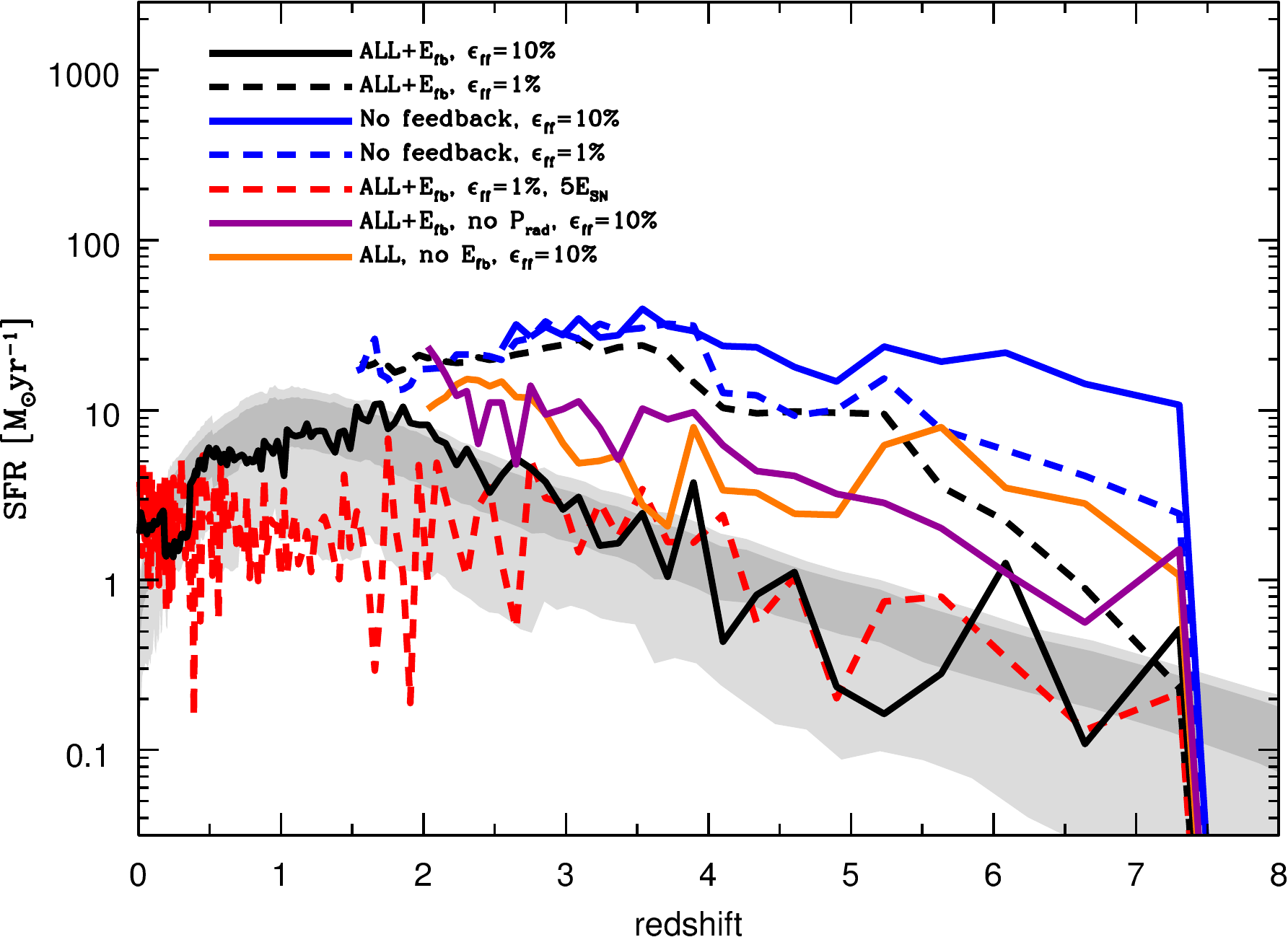}
\caption{Simulated star formation histories compared to the \cite{Behroozi2013} data for $M_{\rm vir}(z=0)=10^{12}\Msol$. Dark and light gray shaded areas are one-and two-sigma confidence regions respectively. We adopt bins of size $\Delta t_{\rm SF}=100\Myr$ for the simulated SFHs. Without feedback, SFRs are overpredicted by at least one order of magnitude at $z>1$. Efficient feedback in conjunction with $\epsilon_{\rm ff}\gtrsim 10\%$ ({\tt ALL\_Efb\_e010}) renders a star formation history in agreement with the Behroozi et al. data. In simulations with a low local star formation efficiency ($\epsilon_{\rm ff}=1\%$), the effectiveness of feedback diminishes and SFRs is $\sim 1$ dex higher than expected. Boosting the available SNe feedback ({\tt ALL\_Efb\_e001\_5ESN}) alleviates this, but leads to a significantly stronger suppression of star formation at $z\lesssim 2.5$. Both radiation pressure and efficient SN feedback appear crucial, as removing any of these feedback sources offsets the SFH by up to $\sim 1$ dex, as discussed in the main text.}
\label{fig:SFH}
\end{center}
\end{figure*}

\begin{figure}[t]
\begin{center}
\includegraphics[scale=0.45]{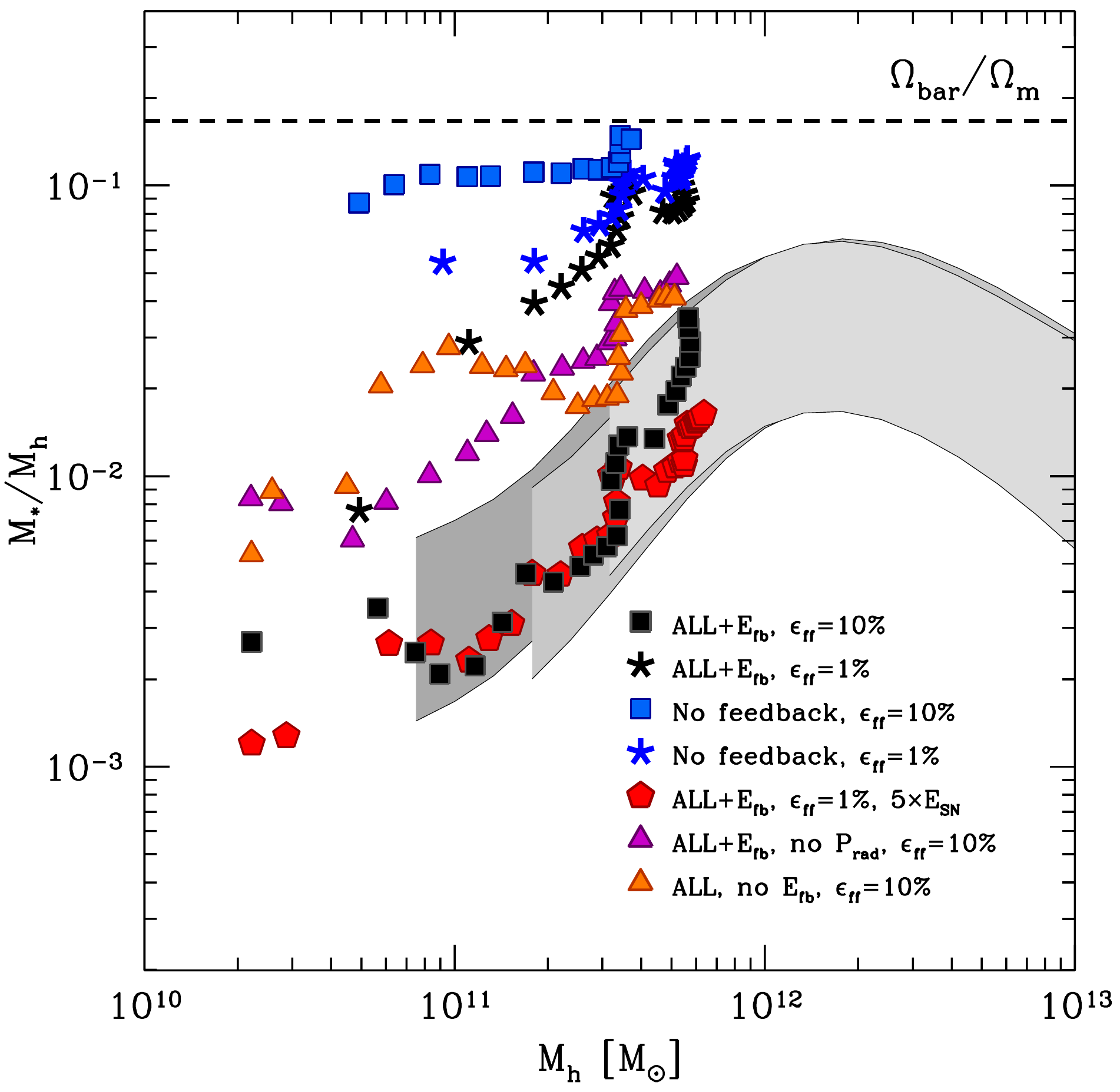}
\caption{The evolution of the stellar mass fraction as a function of halo mass. The shaded regions show, from dark to light grey, the $z=3,2$ and 1 data from \cite{Behroozi2013} where the thickness encompasses $\pm 2\sigma$. The dashed horizontal line show the average cosmic baryon fraction for the adopted cosmology. The simulated data points span the galactic growth from $z=1-7$. As concluded for the star formation histories in Figure \ref{fig:SFH}, the simulations adopting efficient feedback and star formation ({\tt ALL\_Efb\_e010}), as well as boosted SNe feedback ({\tt ALL\_Efb\_e001\_5ESN}) are in good agreement with the semi-empirical relation of Behroozi et al. Removing individual feedback source, or feedback altogether, offsets the simulated data by $\gtrsim1$ dex from the average relation.}
\label{fig:smhm}
\end{center}
\end{figure}
\section{Results}
\label{sect:results}

In this section we present a detailed analysis of a number of basic galaxy properties at $z\gtrsim 1$, relevant for star forming Milky Way analogues, that ought to be reproduced by simulations of galaxy formation: the star formation history, the stellar mass-dark matter halo mass ($M_\star-M_{\rm h}$) relation, the Kennicutt-Schmidt ($\Sigma_{\rm gas}-\Sigma_{\rm SFR}$) relation and the stellar mass-gas metallicity ($M_\star-Z_{\rm gas}$) relation. Furthermore, we study the ability of our models to predict rising or flat rotation curves, a key ingredient in explaining the observed Tully-Fisher relation \citep[galaxy luminosity vs. disk circular velocity, ][]{TF77} for extended spiral galaxies \citep{Reyes2012}.

\subsection{A qualitative comparison}
\label{sect:qualitative}
In Figure \ref{fig:map1} we show large scale maps of the gas surface density, mass weighted temperature, gas metallicity and stellar surface density, at $z=3$, for four of our simulations: {\tt ALL\_Efb\_e010}, {\tt ALL\_Efb\_e001}, {\tt ALL\_Efb\_e001\_5ESN} and {\tt ALL\_e010} (see table\,\ref{table:simsummary1}). From the first two simulations, which \emph{only} differ in their choice of star formation efficiency per free fall time, we find a dramatic difference in outflow properties; for $\epsilon_{\rm ff}=10\%$, galactic winds eject enriched gas from the turbulent galactic disk, while no signs of outflows can be seen when $\epsilon_{\rm ff}=1\%$. In the latter case, almost all metals are retained in the cold star forming gas disk, as is the case for simulations neglecting feedback. The stellar distribution in this simulation is also significantly more compact compared to the other runs.

 Furthermore, the size of the hot gaseous halo surrounding the main progenitor differs between the simulations; in models with inefficient or no feedback, the hot halo forms via cosmological accretion shocks or shocks generated via rapid gravitational potential fluctuations. At $r\sim 50\kpc$, which is close to the virial radius at this redshift, the temperature drops off to $T<10^5\,\K$. In contrast, in simulations with strong feedback-driven winds, the gas outflows contribute significantly to pressurizing the hot halo and driving the outer shock. The hot ($T\gtrsim 10^6\,\K$) halo in such simulations extends far beyond the virial radius of the main dark matter halo.

Boosting the feedback energy per supernova by a factor of five for the case of $\epsilon_{\rm ff}=1\%$ radically changes the mode of galaxy formation, and similar metal enriched outflows and turbulent gas disk morphology as for our fiducial simulations is recovered, at least qualitatively. This shows that there is a certain degeneracy between the star formation efficiency and feedback strength, and a quantitative comparison with observations may be necessary to separate the models. 

We find that neglecting specific sources of stellar feedback leads to significant differences in galaxy evolution. For example, in the 
simulation shown in the rightmost column of Figure \ref{fig:map1} we do not
include the second feedback energy variable, $E_{\rm fb}$, while keeping the rest of the parameters the same as in our fiducial simulation (the leftmost panel). While metal rich outflows are still present, the gaseous disk is significantly less turbulent and is more compact, with less neutral gas extending to large distances ($\sim 10$ kpc in the fiducial run), as seen in the temperature map. This results in a more massive stellar system, which as we demonstrate below is in tension with semi-empirically derived stellar mass-halo mass relations \citep{Behroozi2013}. A similar conclusion holds for the simulations that neglect radiation pressure.

\subsection{Star formation histories}
\label{sect:SFH}
Figure \ref{fig:SFH} shows the star formation histories (SFHs), calculated in bins of $\Delta t=100\Myr$\footnote{Note that the degree of fluctuations in star formation rate is sensitive to the choice of $\Delta t_{\rm SF}$ and can vary with stellar mass, as reported by \cite{Hopkins2014} \citep[see also][]{Feldmann2012}. The scatter in our simulated galaxy increases towards higher redshift as star formation is found to be highly episodic in the low mass progenitors ($M_{\rm vir}<10^{11}\Msol$). }, for the simulated galaxies compared to the semi-empirically inferred SFH from \cite{Behroozi2013} relevant for a galaxy forming in a $M_{\rm vir}(z=0)=10^{12}\Msol$ dark matter halo. Regardless of the choice of star formation efficiency per free fall time, neglecting feedback leads to a dramatic overestimate of the galactic SFR at all redshifts by $\gtrsim 1\,{\rm dex}$ compared to the predictions by Behroozi et al.. This may seem counterintuitive as the lower abundance of H$_2$ in dwarf galaxies at high redshifts is thought to make star formation less efficient. However, as the ISM self-enriches via SNe, and no stellar feedback is present to drive metal rich winds, a larger fraction of the gas mass rapidly becomes available for star formation due to the effectively lower density threshold via the higher $f_\HH$, see \S\,\ref{sect:h2model}. As mentioned above, this is the case regardless of the adopted value for $\epsilon_{\rm ff}$, although the normalization of the relation at $z>4$, and hence how rapidly the galaxy self-enriches, depends on the precise value. 

Simply incorporating efficient stellar feedback (\S\,\ref{sect:FB}) in the KMT09 model does not necessarily overcome this problem. The simulation with $\epsilon_{\rm ff}=1\%$ overpredicts the SFRs by up to a factor of ten and the star formation rate in this case is not significantly affected by feedback. For star formation to be sufficiently feedback regulated, the local star formation efficiency per free fall time needs to be sufficiently large, here $\epsilon_{\rm ff}=10\%$. Once this is satisfied, the simulations are in excellent agreement with the data of \cite{Behroozi2013}. 

In a star formation model based on the abundance of H$_2$, such as KMT09, the gas metallicity plays an important role in setting the fraction of gas available for star formation. The local metallicity, in turn, is regulated by the feedback driven  outflows. In our current simulation suite, this only occurs if star formation, and hence feedback, becomes sufficiently spatially and temporally correlated. As we show in \S\,\ref{sect:massmet}, the simulations with efficient wind driving also match the observed evolution of the relation between stellar mass and gas metallicity. 
 
Star formation suppression can also be achieved by increasing the available SN thermal energy budget, here illustrated by employing a boost by a factor of five for the run with $\epsilon_{\rm ff}=1\%$. The resulting SFH agrees almost perfectly with the less energetic, but self-regulated, fiducial simulation at $z\gtrsim 3$. As discussed in \S\,\ref{sect:qualitative}, this illustrates a certain degeneracy between detailes of star formation and feedback prescriptions in such simulations, which needs to be broken by other observables, especially because the feedback boosted simulation severely distorts the gas disk at $z<2$, as seen in Figure \ref{fig:map2}.

In Figure \ref{fig:SFH} we also show the impact of neglecting various sources of stellar feedback in our fiducial simulation. By not considering radiation pressure feedback, star formation rates increase by a factor of several at all redshifts, as found in \cite{Agertz2013} for isolated disks. Reducing the efficiency of thermal feedback by neglecting the feedback energy variable significantly increases the SFRs at $z>4$, while bringing them into agreement with the Behroozi data at later times. This behavior stems from the inability of radiative feedback to efficiently regulate star formation in low metallicity gas at high redshifts, as photon trapping via dust becomes negligible, whereas this is not the case in the more enriched disk at late times. This collective, and highly non-linear behavior of early radiative feedback and SNe, was recently studied in a fully cosmological setting by \cite{Hopkins2014} who also found that it was necessary to consider these two feedback processes jointly in order to reproduce observationally derived star formation histories.

\begin{figure*}[t]
\begin{center}
\begin{tabular}{cc}
\includegraphics[scale=0.4]{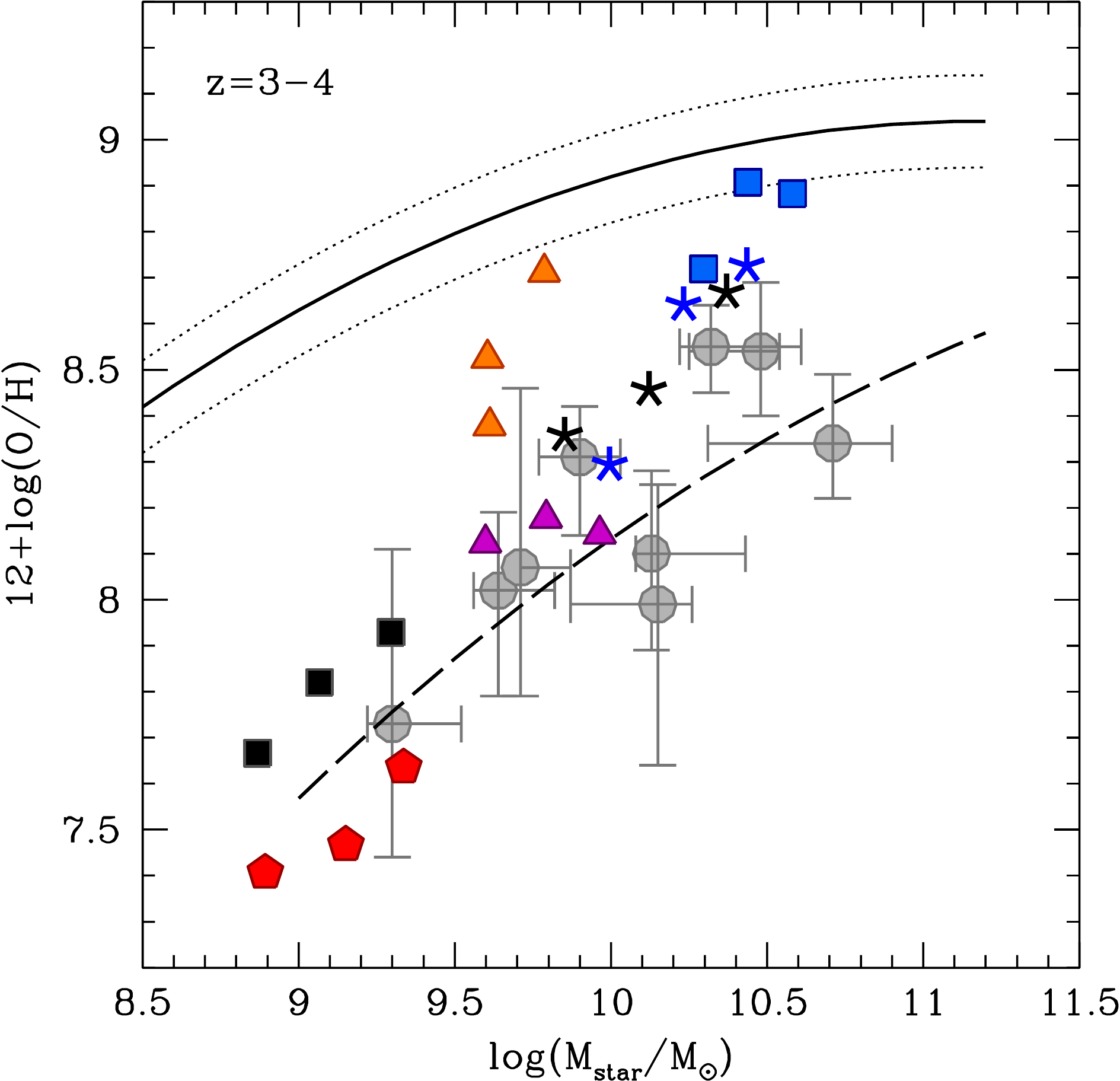}
\includegraphics[scale=0.4]{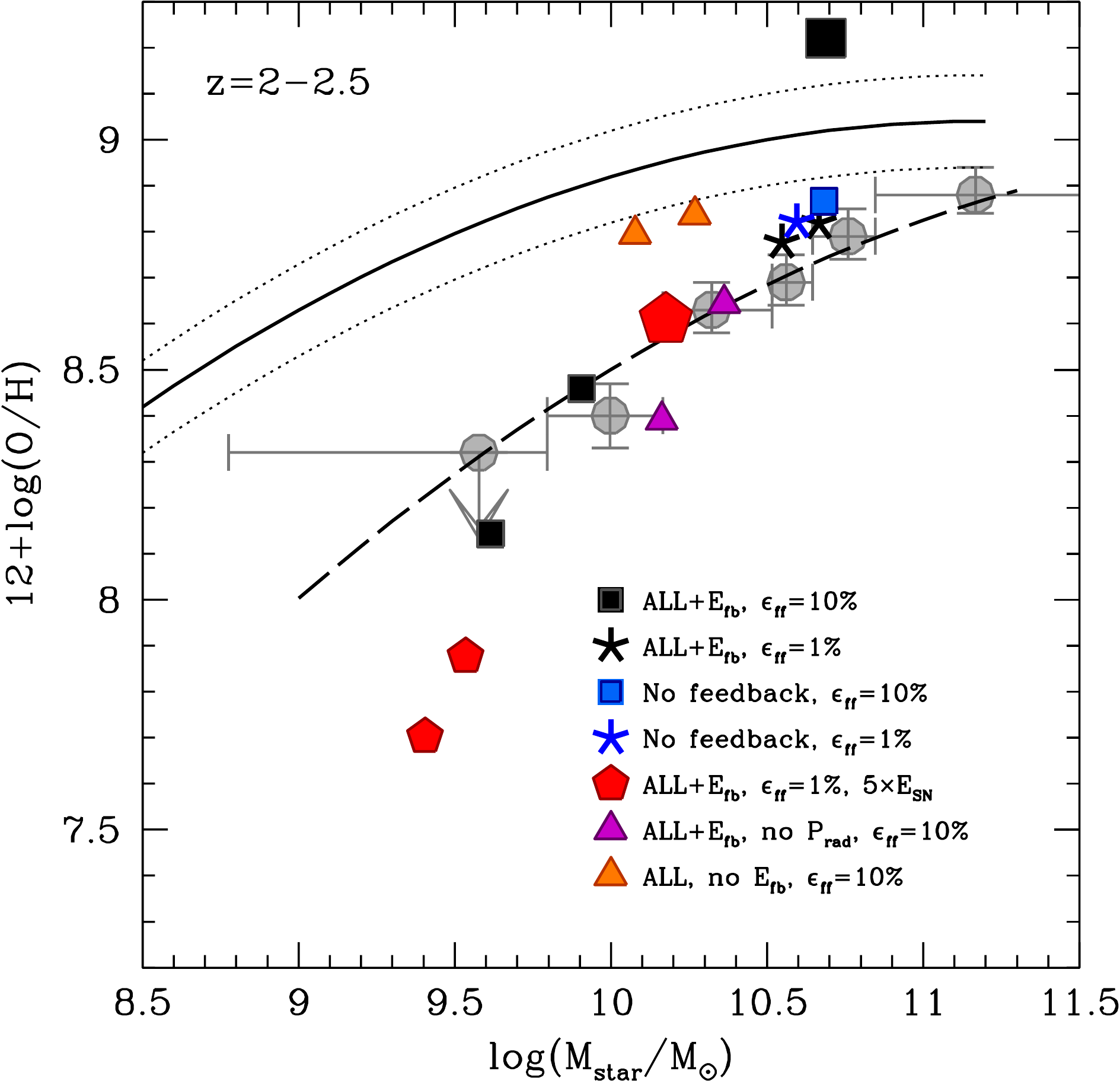}
\end{tabular}
\caption{The stellar mass--cold gas metallicity relation at $z=3-4$ (left) and $z=2-2.5$ (right). The upper solid line shows the relation and its dispersion observed at $z\sim0.07$, as inferred by \cite{KewleyEllison2008}. (Left) Grey symbols show observational data from \cite{Maiolino2008}, adopting the same metallicity calibration as \cite{KewleyEllison2008}, for individual galaxies at $z\sim 3.5$ where the lower dashed line is a fit to the data. When star formation is feedback regulated, the simulations conform with the observed $M_\star-Z_{\rm gas}$ relation. Without any feedback, or in the case of low star formation efficiency ({\tt ALL\_ Efb\_e001}), the galaxy rapidly evolves to a relation more akin to what is observed for $z\sim 0$ galaxies. When the available supernovae feedback energy is boosted by a factor of 5, the metal content stays lower than for the other runs (for the same stellar mass). (Right) Grey symbols here show observational data from \cite{Erb2006}, using the same calibration as the above data sets \citep[see][]{Maiolino2008}, for galaxies at $z=2.26\pm 0.17$. Simulations without efficient star formation regulation is here no longer in disagreement with the observations, with the different models evolving ``along'' the relation, making the role of metal rich outflows in setting the normalization of the $M_\star-Z_{\rm gas}$ relation less obvious \citep[see also][]{Tassis08}. The large points show the $z=0$ results for {\tt ALL\_Efb\_e001\_5ESN} and {\tt ALL\_Efb\_e010}.
}
\label{fig:MZ}
\end{center}
\end{figure*}

\subsection{The stellar mass-halo mass relation}
\label{sect:smhm}
In Figure \ref{fig:smhm} we show the stellar mass fraction ($M_\star/M_{\rm h}$) vs. halo mass relation for the simulated galaxies. The shaded regions show the inferred $2\sigma$ relations for $z=3,2,1$, from \cite{Behroozi2013}.  For consistency with \cite{Behroozi2013}, we use the virial mass definition of \cite{BryanNorman1998} to define the halo mass of the progenitor. The $M_\star/M_{\rm h}$ evolutionary tracks are shown for all simulations at $z\gtrsim1$, wherever simulation data exists. We note that the $M_\star/M_{\rm h}$ relation on occasion rapidly evolves vertically, or that $M_{\rm h}$ even decreases temporarily. This behavior  stems from major merger events which not only boosts star formation, but can complicate measurements of the halo virial mass.

Note that the $M_\star/M_{\rm h}$ relation and SFHs in the previous section are not independent constraints. Indeed, simulations that also match the inferred SFHs in the previous section are in good agreement with the predicted stellar mass fractions, i.e. runs employing $\epsilon_{\rm ff}= 10\%$ and/or efficient feedback ({\tt ALL\_Efb\_e010} and {\tt ALL\_Efb\_e001\_5ESN}). Inefficient local star formation ($\epsilon_{\rm ff}=1\%$) overpredicts the stellar content by an order of magnitude, while in runs in which  feedback is neglected the stellar fraction is close to the mean cosmic baryon fraction at all times. 

The interplay between radiation pressure and efficient thermal feedback in establishing a realistic stellar mass fraction is illustrated whenever either one of these sources is removed from the feedback budget; the stellar fraction is suppressed to a much greater degree at late times (i.e. more massive dark matter halos) when $E_{\rm fb}$ is neglected, and the opposite is true when radiation pressure is neglected.

\subsection{The mass-metallicity relation and effective yields}
\label{sect:massmet}
Figure \ref{fig:MZ} shows the stellar mass-gas metallicity ($M_\star-Z_{\rm gas}$) relation for the simulated galaxies at $=2-2.5$ and $z=3-4$. Note that the gas metallicity plotted in the figure is measured for the cold gas component of the galaxies. We compare the simulations with observational data of galaxies at $z\sim0.07$ \citep[as inferred by][]{KewleyEllison2008}, $z\sim2.2$ \citep{Erb2006} and $z\sim3.5$, where a uniform calibration of metallicity indicators was used across all redshifts \citep{Maiolino2008}. In order to compare the data to the observational aperture adopted by \cite{Maiolino2008}, we quantify the gas metallicity as the mass weighted mean metallicity at radii $r\leq3$ kpc. The stellar mass is the total stellar mass for $r\leq10\kpc$, which safely contains all stellar mass belonging to the central galaxies at all redshifts under investigation. As we only track the \emph{average} metallicity of the gas in {\tt RAMSES}, we calculate $12+\log({\rm O/H})$ assuming solar mixture and adopt $12+\log({\rm O/H})_\odot=8.69$ for the solar value \citep{Asplund2009}.

From Figure \ref{fig:MZ} we find that \emph{not} matching the SFH, $M_\star/M_{\rm halo}$ and KS relations in the previous sections may still allow the galaxy to conform to the observed $M_\star-Z_{\rm gas}$ relation at $z>2-3$. The fact that the $M_\star-Z$ relation is determined primarily by the overall efficiency of galactic star formation, and not necessarily via properties of feedback-driven outflows, has been emphasized before \citep{Brooks07,Tassis08}. The almost 2 dex spread in stellar mass ($8.5<M_\star<10.5$) in the simulation suite measured at $z=3-4$ forms a steeper linear relation, $12+\log({\rm O/H})\approx 7.5 + \log(M_\star/10^9 M_\odot)$. Individual simulations, e.g. the favoured {\tt ALL\_Efb\_e010} run, trace a more shallow relation over the same redshift range. When $\epsilon_{\rm ff}=1\%$ ({\tt ALL\_ Efb\_e001}), or no feedback is present, star formation is not efficiently regulated, leading to increasing stellar masses and metallicities that eventually causes the galactic average to diverge from the mean relation at lower redshifts. However, as the $z\sim2-2.5$ data \citep{Erb2006} form a steeper relation than that at $z\sim0$, meaning the metallicities at the high stellar mass end show a weaker evolution, these particular simulations are not in strong disagreement with observations below $z\sim2$. 

The fiducial simulation ({\tt ALL\_ Efb\_e010}) is in excellent agreement with observations at all times. In the case of boosted SNe feedback energy ({\tt ALL\_Efb\_e001\_5ESN}), gas metallicities are lower at all redshifts, possibly in tension with observed gas metallicities at $z\sim 2-2.5$, although at the low stellar masses under consideration ($M_\star\sim3\times 10^{9}\,\Msol$) the metallicity measurements are only upper limits. At $z=0$, the fiducial model shows a high central metallicity, but is still in broad agreement with observations, while the boosted feedback energy model is metal deficient and lies close to the high redshift ($z\sim2-2.5$) relation.

Even though neglecting radiation pressure overestimates stellar masses, see \S\,\ref{sect:smhm}, the enrichment history allows the galaxy to evolve ``along'' the evolving $M_\star-Z_{\rm gas}$ relation. This is not the case when the second feedback energy variable $E_{\rm fb}$ is neglected, as the metal rich gas disk can be seen to evolve off the observed relation already at $z\sim 4$, illustrating the need in our current models for efficient thermal feedback to regulate the galactic metal content, at least at this specific epoch.

\begin{figure}[th]
\begin{center}
\includegraphics[scale=0.4]{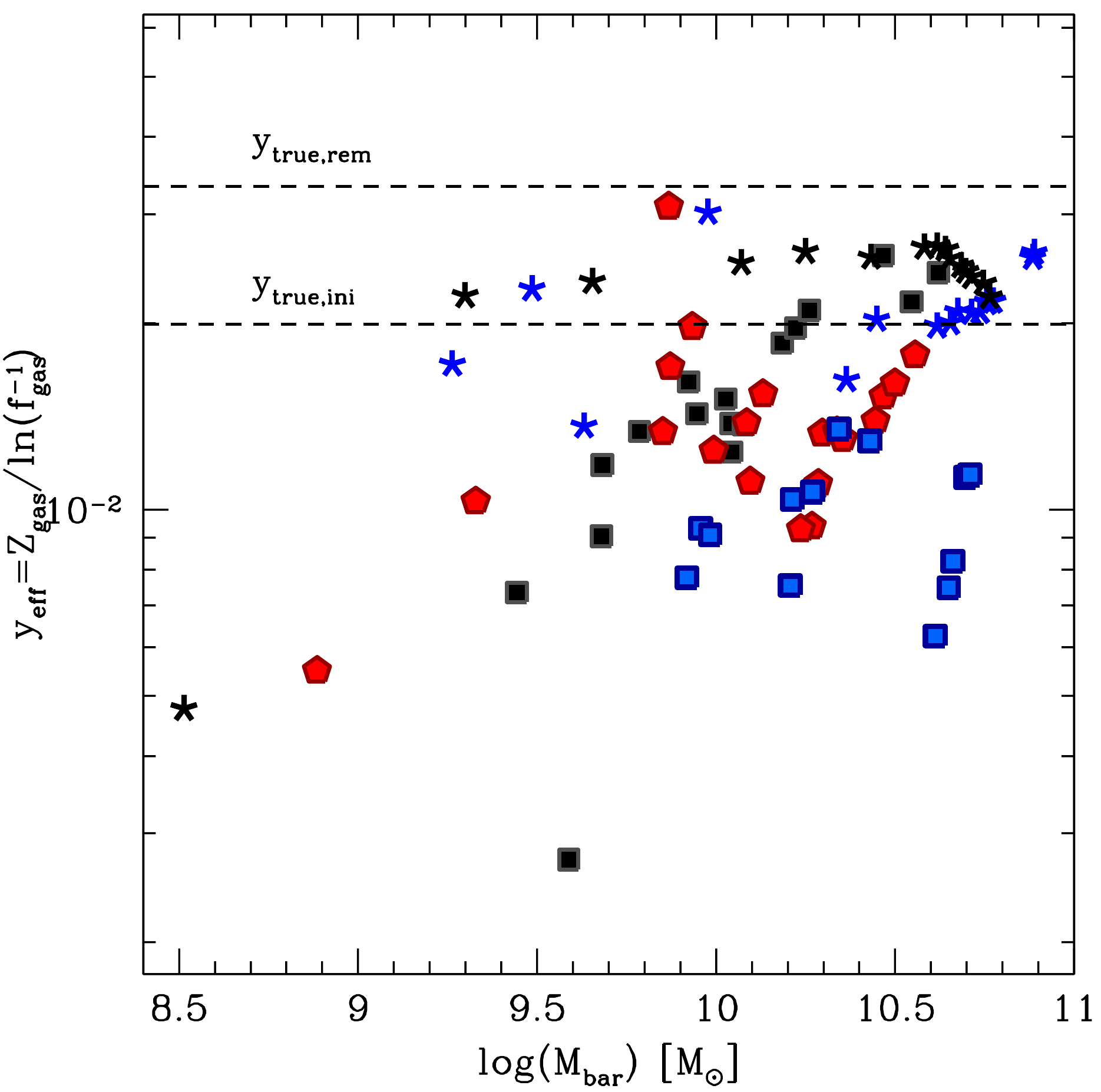}
\caption{Effective yields for the main progenitor at $z=1-7$ in the simulation suite. The data points are the same as in Figure \ref{fig:MZ}. A global trend is found for all galaxies to have yields lower than the expected true yields from a closed box model, but not necessarily for the same reason. The fiducial {\tt ALL\_Efb\_e010} simulation shows a clear evolution from low yields ($y_{\rm eff}<10^{-2}$), due to metal rich outflows, to a state where $y_{\rm eff}\sim y_{\rm true}$ for $M_{\rm bar}\gtrsim 10^{10}\,M_\odot$. We refer to the main text for a detailed discussion.
}
\label{fig:yield}
\end{center}
\end{figure}

\begin{figure*}[th]
\begin{center}
\begin{tabular}{ccc}
\includegraphics[scale=0.31]{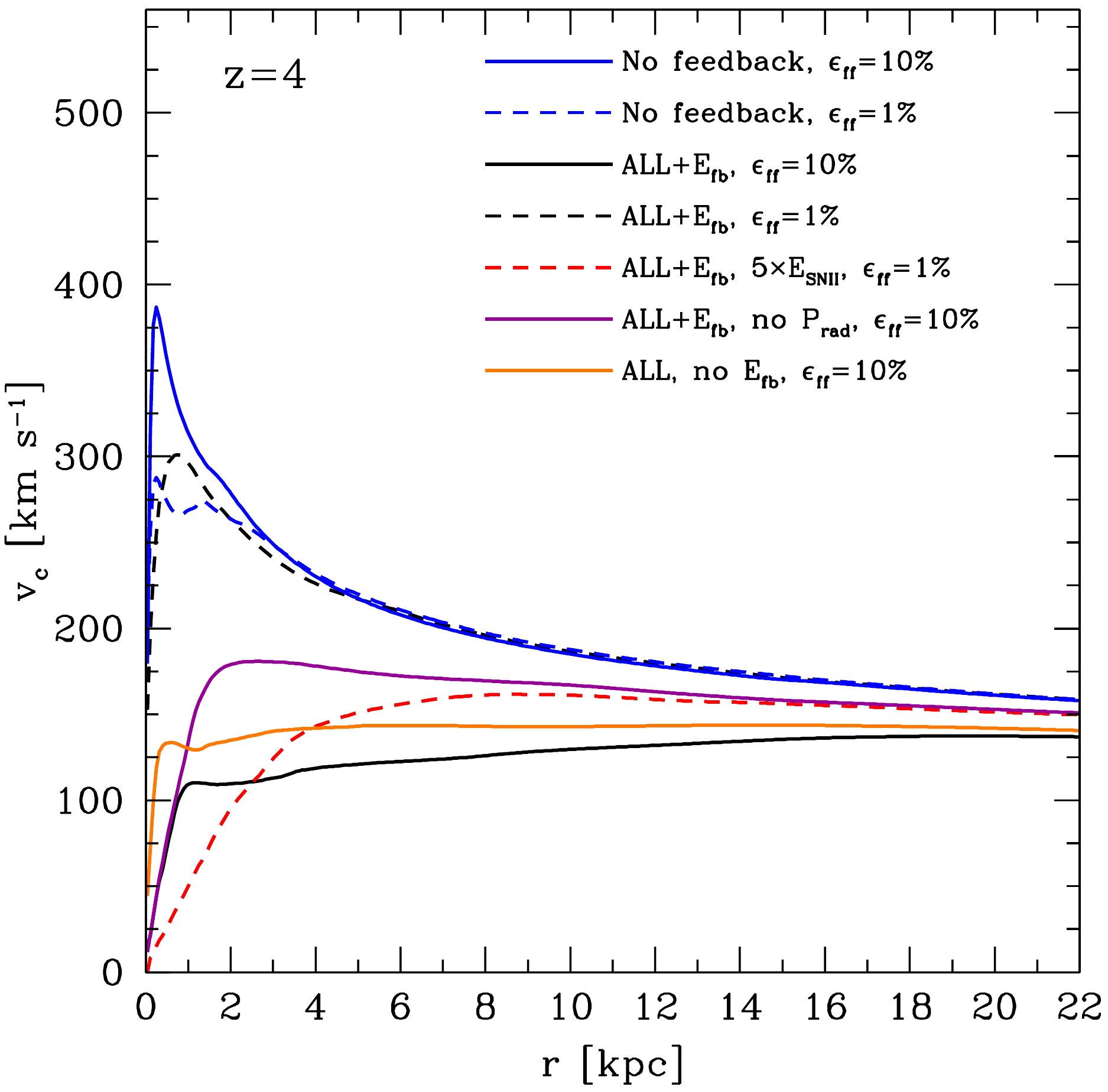}
\includegraphics[scale=0.31]{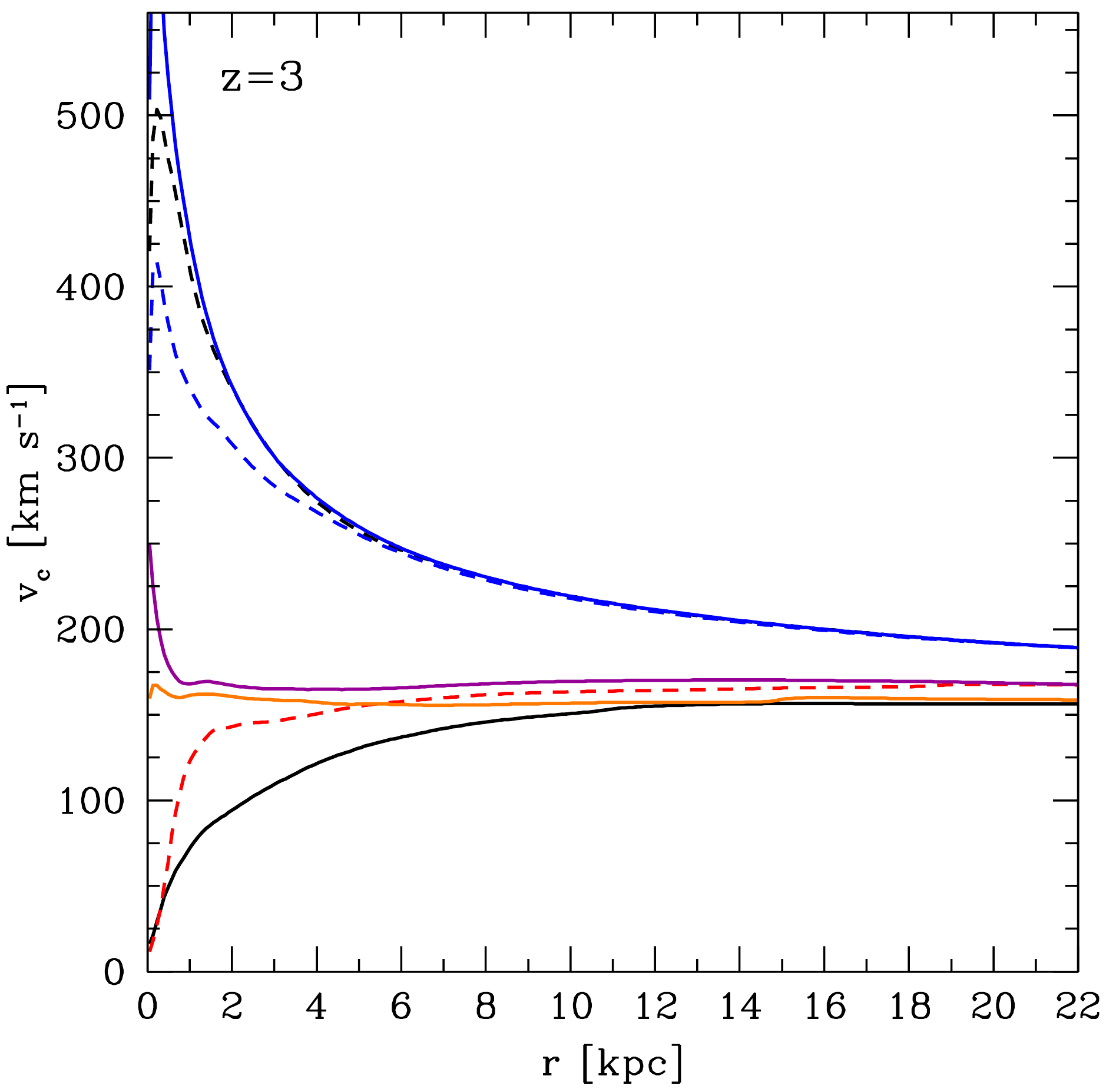}
\includegraphics[scale=0.31]{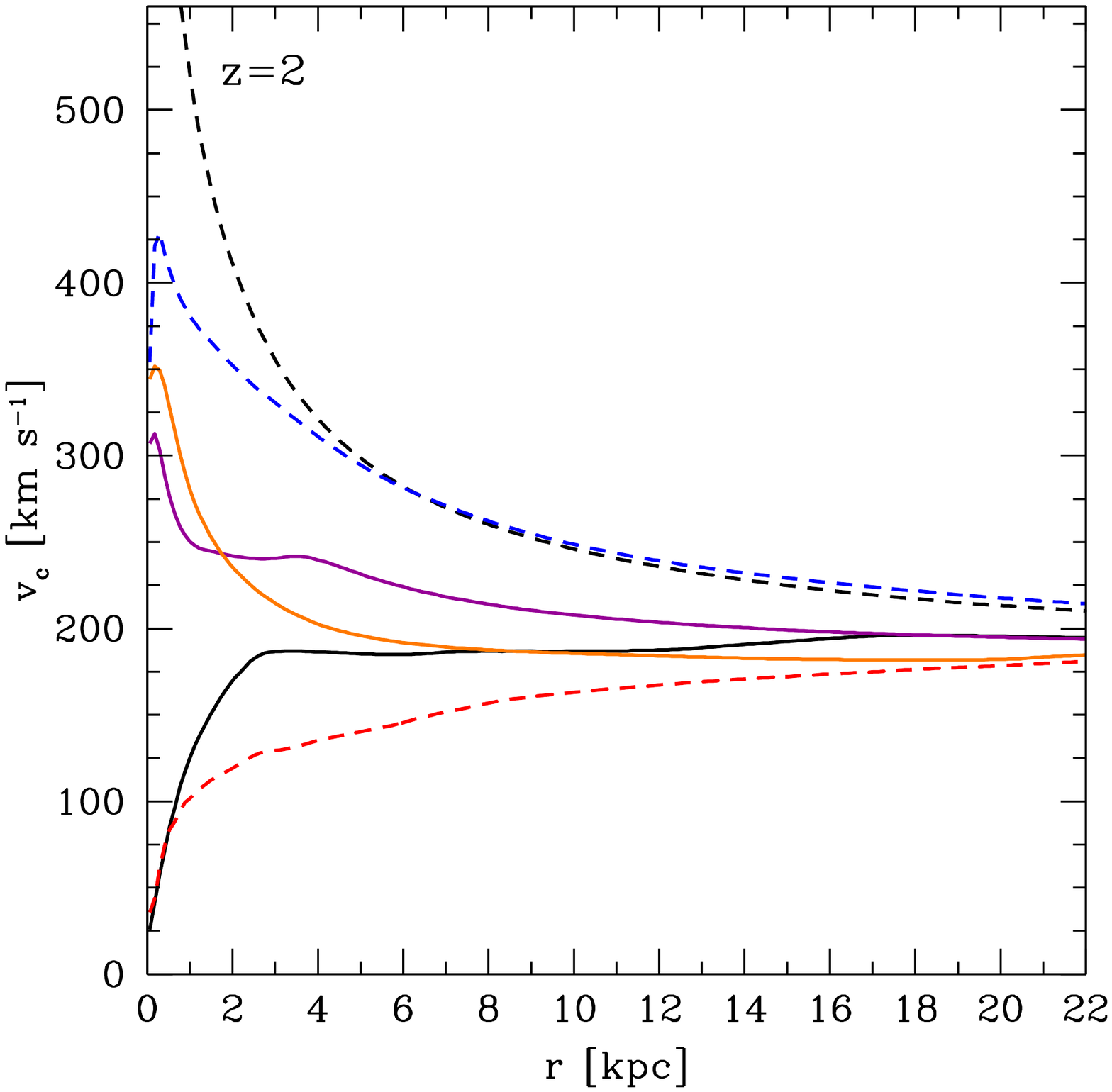}
\end{tabular}
\caption{Circular velocities for the entire simulation suite at $z=4$ (left), $z=3$ (middle) and $z=2$ (right). The only two simulations that maintain a rising or flat circular velocity profile are the fiducial simulation ({\tt ALL\_Efb\_e010}) and the boosted SNe feedback run ({\tt ALL\_Efb\_e001\_5ESN}). As argued in the main text, only these two simulations regulate the star formation rates to reasonable levels while driving galactic winds. When radiation pressure or efficient thermal feedback is removed, the resulting rotation curves remain flat until $z\sim 3$, after which inefficient removal of low angular momentum material leads to a significant upturn in circular velocities in the central parts of the galaxies.
}
\label{fig:vc}
\end{center}
\end{figure*}

\begin{figure}
\begin{center}
\includegraphics[width=0.475\textwidth]{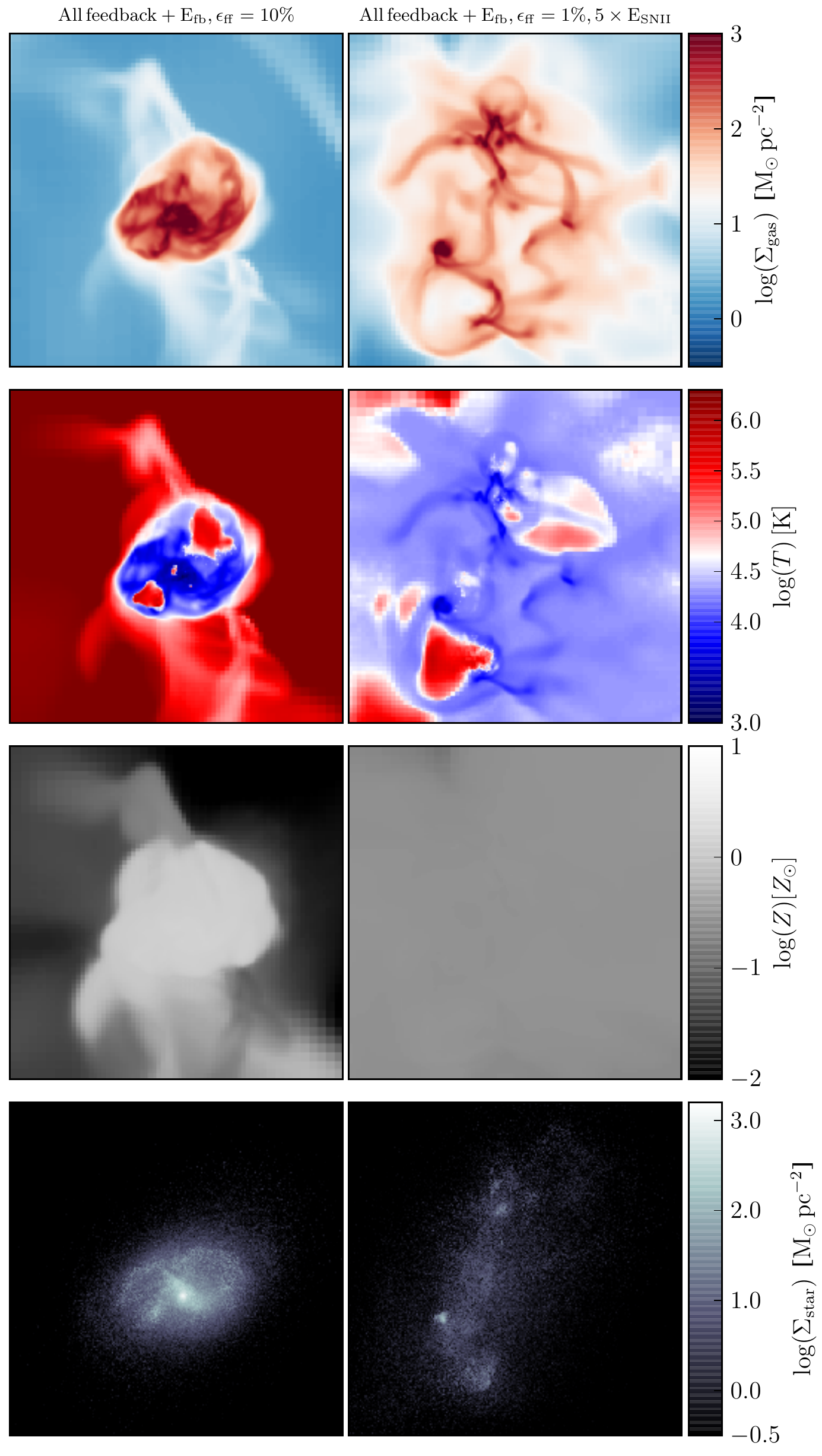}
\caption{Maps of, from top to bottom, gas surface density, mass weighted temperature, gas metallicity and stellar surface density stellar for, from left to right, the fiducial {\tt ALL\_Efb\_e010} run and {\tt ALL\_Efb\_e001\_5ESN}, both at $z=1$. The maps span 30 kpc on each side. As detailed in the text, both simulations compare favorably to the observed $\Sigma_{\rm gas}-\Sigma_{\rm SFR}$ relation, the stellar mass-gas metallicity relation ($M_\star-Z_{\rm gas}$), and the stellar mass-dark matter halo mass ($M_\star-M_{\rm vir}$) relation. However, the galactic morphologies are dramatically different; whereas the galaxy in fiducial simulation run eventually settles into an extended disk at the peak of star formation ($1<z<2$), the simulation adopting boosted SNe feedback, and a low star formation efficiency per free fall time, produces a heavily distorted gas distribution with a dispersion dominated stellar component.}
\label{fig:map2}
\end{center}
\end{figure}

\subsubsection{Effective yields}
To better understand the role of metal rich outflows, we calculate effective yield, defined as
\begin{equation}
\label{eq:yield}
y_{\rm eff}=\frac{Z_{\rm gas}}{\ln(1/f_{\rm gas})},
\end{equation}
where $f_{\rm gas}= M_{\rm gas}/(M_{\rm gas}+M_\star)$ is the fraction of baryons in the gas phase. The effective yield has been widely used as a diagnostic of the evolution of the baryonic component of galaxies, and more specifically as a test of the validity of the closed-box approximation \citep{PagelPatchett1975,Edmunds1990}. Observationally, the effective yield is known to decrease with galactic mass \citep{Tremonti2004}, with a sharp decline around the mass of dwarf galaxies \citep[$\vel_{\rm rot}\lesssim 100\,{\rm km}\,{\rm s}^{-1}$,][]{Garnett02}.

Under the closed-box assumption, the effective yield is always equal to the true yield $y_{\rm true}$,  typically defined, for a single stellar population, as the mass in newly synthesized metals returned to the ISM normalized to the stellar mass of this population locked up in stellar remnants and long-lived stars, i.e.
\begin{equation}
y_{\rm true,rem}=\frac{1}{1-R}\int_{0.1}^{100} m p_{\rm im}\phi(m)dm,
\end{equation}
where $m$ is the stellar mass, $\phi(m)$ the IMF, $p_{\rm im}$ is the instantaneous stellar yield, and $R$ the mass fraction returned to the ISM. For our feedback prescription we calculate the true yield, as well as the initial true yield, $y_{\rm true,ini}$, where we consider $R=0$, i.e. the stellar population is assumed to retain all of its initial birth mass.

Following \cite{Tassis08}, we calculate the ``observed'' effective yield as a function of total baryons mass in our main galaxy using Equation\,\ref{eq:yield}, where we consider only the cold gas ($T\le10^{4}$ K) and the metal content within the stellar extent defined as the radius that includes 90\% of the total stellar mass. The result for the entire simulations suite, over the redshift range $z=1-7$, is presented in Figure \ref{fig:yield}. Our fiducial simulation ({\tt ALL\_Efb\_e010}) shows a clear plateau towards the true yield for $M_{\rm bar}\gtrsim 10^{10}M_\odot$, with lower yields signifying outflows \citep[see also][]{Brooks07}, as seen in the metal rich winds in Figure \ref{fig:map1}. The same is true for the feedback boosted simulations, although the average values for $M_{\rm bar}\gtrsim 10^{10} M_\odot$ are lower than for the fiducial case, indicating that outflows are still prominent.

Figure \ref{fig:yield} shows that simulations with a low star formation efficiency, {\tt ALL\_Efb\_e001} and {\tt NoFB\_e001}, have effective yields close to the true yield already at yearly times, although the former simulation is significantly offset at $z\gtrsim6$ ($\log (M_{\rm bar})=8.5$), indicating that the ``observed'' effective yield may not solely be explained via galactic winds \citep{Tassis08}. 

A peculiar result is found for the simulation without feedback and $\epsilon_{\rm ff}=10\%$ ({\tt NoFB\_e010}), which is shown for $2.5<z<7$. The effective yields lie significantly below all other data for $M_{\rm bar}\gtrsim 10^{10} M_\odot$, despite having no means of ejecting enriched gas. The reason for this is that, despite being enriched to $Z_{\rm gas}>Z_\odot$ already at early times, the gas fraction is kept very low due to the short depletion time-scale and low effective density threshold for star formation via the KMT09 model. In {\tt ALL\_Efb\_e010}, $f_{\rm gas}(r<10\kpc)\sim65\%$ at $z=2.5$, compared to $f_{\rm gas}\sim 9\%$ for {\tt NoFB\_e010}, which pushes $y_{\rm eff}$ to lower values in the latter simulation.

\subsection{Circular velocity profiles}

Simulated galaxies have traditionally displayed high central concentrations of baryons, leading to strongly peaked circular velocities towards the galactic center \citep{NavarroWhite93,Abadi03b,Okamoto05,Scannapieco09,Piontek09b,Hummels2012}. Removal of preferentially \emph{low} angular momentum gas via stellar feedback can remedy this problem \citep[e.g.][]{Governato07,Agertz2011,Brook2012,Ubler2014}, hence bringing simulated galaxies closer to the observed Tully-Fisher relation \citep{TF77,Pizagno07}. 

Figure \ref{fig:vc} shows circular velocities of the entire suite of models, adopting the KMT09 model, at $z=4$, 3 and 2. Strongly peaked circular velocities are found already at $z=4$ for all models with no feedback, or where the star formation efficiency is too low ($\epsilon_{\rm ff}=1\%$) to allow for efficient wind driving. Adopting efficient feedback and efficient local star formation, as in our fiducial run ({\tt ALL\_ Efb\_e010}), as well as boosting the available SNe energy ({\tt ALL\_Efb\_e001\_5ESN}), lead to a rising or flat rotation curve at all times due to efficient removal of low angular momentum gas in feedback driven outflows. Neglecting either radiation pressure or efficient thermal feedback via $E_{\rm fb}$ results in a massive bulge component at $z\lesssim 2$, demonstrating their important interplay in establishing galactic properties.

\begin{figure*}[t]
\begin{center}
\begin{tabular}{cc}
\includegraphics[scale=0.4]{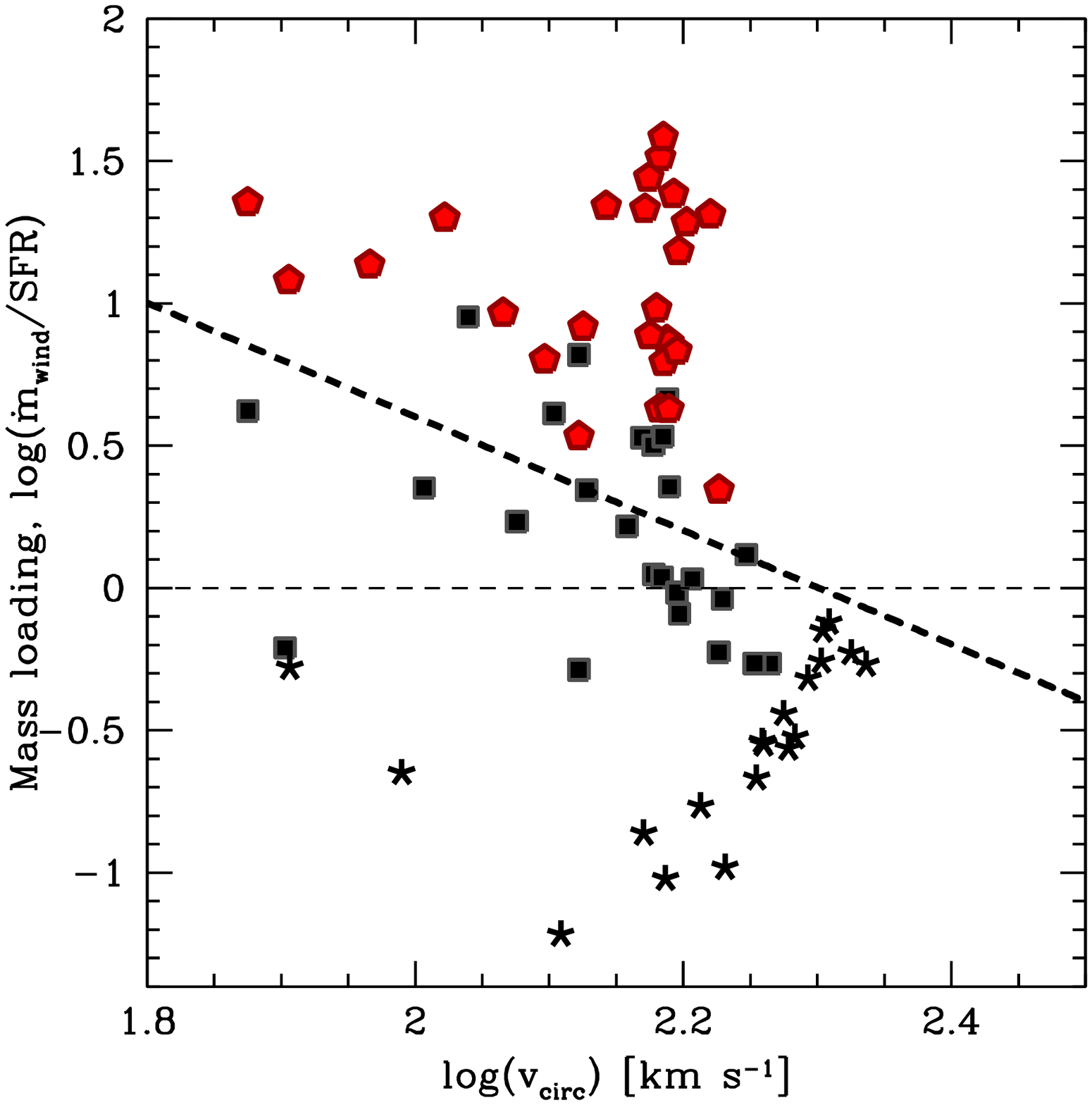}
\includegraphics[scale=0.4]{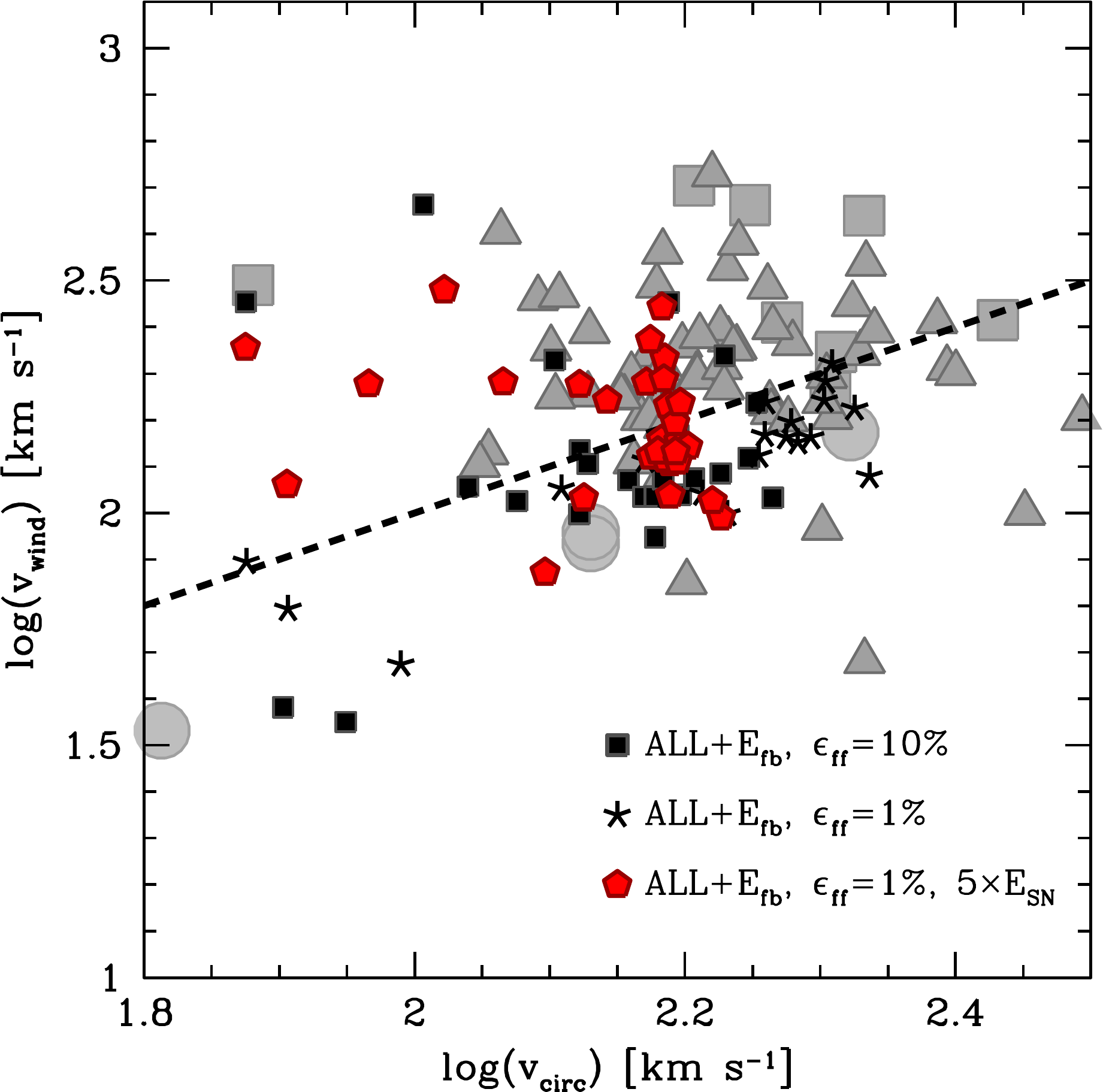}
\end{tabular}
\caption{(Left) Mass loading factors ($\eta$) of galactic outflows and (right) average wind velocity as a function of circular velocity at $r=20\kpc$ for {\tt ALL\_Efb\_e001} (black stars), {\tt ALL\_Efb\_e010} (filled black squares) and {\tt ALL\_Efb\_e001\_5ESN} (filled red pentagons). See the main text for details of how the quantities are computed. The wind mass loading factors differ significantly between the simulations; {\tt ALL\_Efb\_e001} fails to drive winds, with mass loading factors in excess of unity for any $\vel_{\rm circ}$, whereas the fiducial {\tt ALL\_Efb\_e010} produces increasing $\eta$ with decreasing $\vel_{\rm circ}$. In contrast, mass loadings in {\tt ALL\_Efb\_e001\_5ESN} are large ($\eta\gtrsim 10$) for all circular velocities. For comparison, the dashed line shows $\eta\propto \vel_{\rm circ}^{-2}$, relevant for energy driven outflows \citep[e.g.][]{oppenheimerdave06}. Despite these significant differences, the wind velocities are roughly the same in all cases (right), and broadly agrees with observations. The grey filled circles and squares show observed outflow velocities inferred from Na\,I absorption features in starburst galaxies, adapted from \cite{SchwartzMartin2004} and \cite{Rupke2005b} respectively. The filled grey triangles show outflow velocities derived from Mg\,II absorption lines from a sample of galaxies at $0.3<z<1.4$ from \cite{Rubin2014} (here the central velocity, $\vel_{\rm flow}$, in the two-component model in Rubin et al.). The dashed line indicates $\vel_{\rm wind}=\vel_{\rm circ}$.
}
\label{fig:wind}
\end{center}
\end{figure*}

\begin{figure*}[th]
\begin{center}
\includegraphics[width=0.85\textwidth]{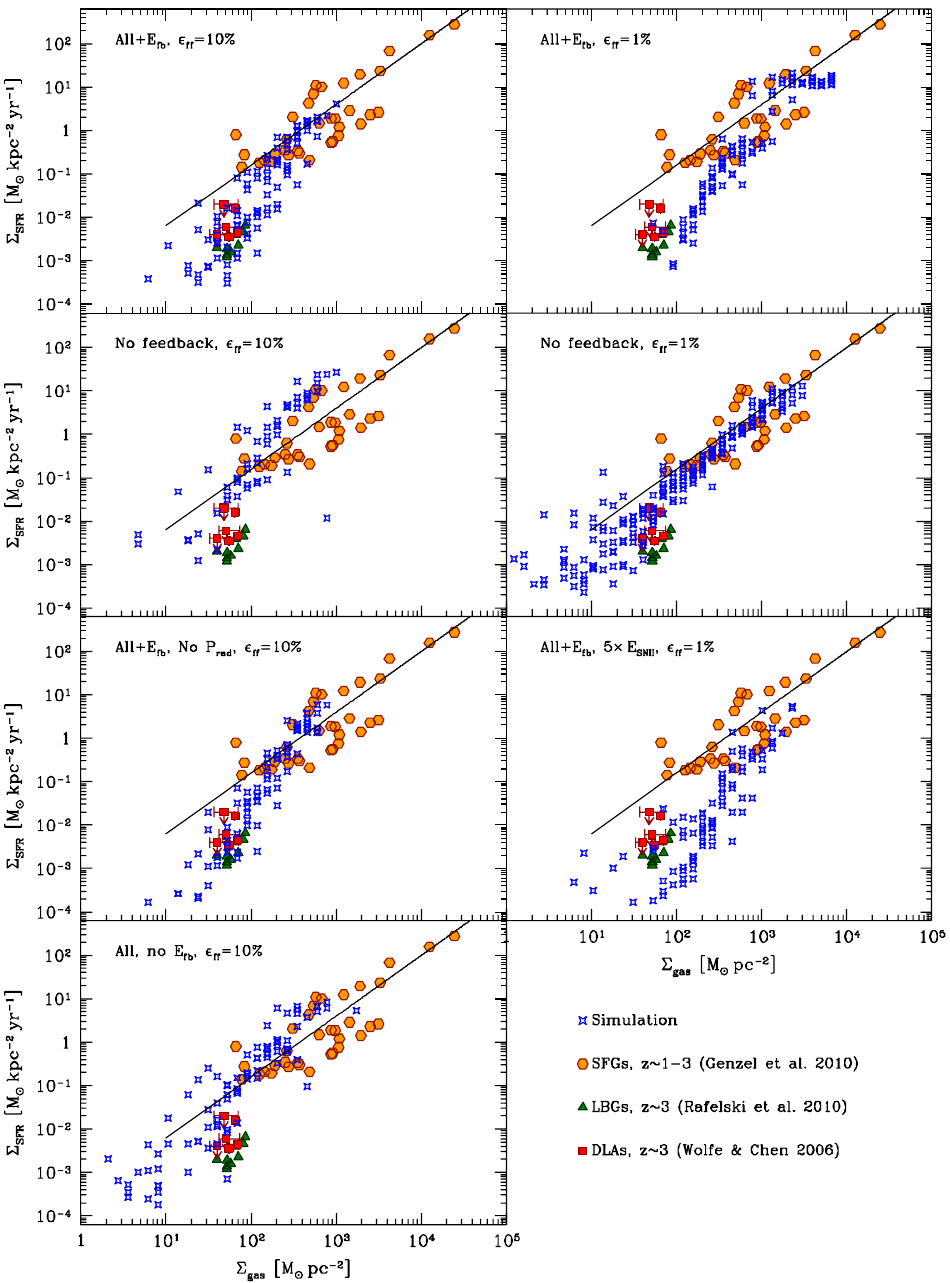}
\caption{The Kennicutt-Schmidt (KS) relation ($\Sigma_{\rm gas}-\Sigma_{\rm SFR}$) at $z=2-3$. Relation from the simulated galaxies (blue crosses) is compared to relations derived from observations: CO data for normal star forming galaxies at $z\sim 1-3$ \citep[orange hexagons,][]{Genzel2010}, DLAs at $z\sim3$ \citep[red squares,][]{WolfeChen06} and LBGs at $z\sim 3$ \citep[dark green circles,][]{Rafelski2011}. The solid black line represents the average $z=0$ relation of \cite{kennicutt98}, i.e. $\Sigma_{\rm SFR}=2.5\times 10^{-4}\Sigma_{\rm gas}^{1.4}\,\Msol{\rm kpc}^{-2}{\rm yr}^{-1}$. The star formation-feedback interplay establishes a wide array of relations, where the best match to observations is found for the fiducial simulation ({\tt ALL\_Efb\_e010}) which predicts the normalization of the KS relation for $\Sigma_{\rm gas}\gtrsim 10^2\,\Msol{\rm pc}^{-2}$ as well as the truncation at lower surface densities, in agreement with the DLA/LBG observations. See main text for a discussion of the entire simulations suite. }
\label{fig:KS1}
\end{center}
\end{figure*}

\section{Breaking the degeneracy}
\label{sect:degeneracy}

As we have demonstrated in the previous section, the best matches to observations are found for two of our models: {\tt ALL\_Efb\_e010} and {\tt ALL\_Efb\_e001\_5ESN}. Both models of galaxy formation are able to reproduce global galactic characteristics, despite the significantly different star efficiency values and amount of feedback energy. To break this degeneracy, other properties of simulated galaxies need to be considered. Potentially, this can include a variety of observations, including studies of the circum-galactic medium (CGM), absorption lines studies of multiphase gas in the galactic halo, detailed properties of the stellar disks etc. For now, we will study the morphological state of the main galaxy progenitor at lower redshifts, properties of galactic winds, as well as internal star formation properties, here the $\Sigma_{\rm gas}-\Sigma_{\rm SFR}$ (Kennicutt-Schmidt) relation.

\subsection{Morphology}
\label{sect:morphology}
Figure \ref{fig:map2} shows the main galaxy progenitors in {\tt ALL\_Efb\_e010} and {\tt ALL\_Efb\_e001\_5ESN} at $z=1$. At his epoch, the fiducial simulation has transitioned into a quiescent state of star formation, with a prominent (turbulent) gaseous disk in place, as well as an emerging stellar disk where stars form in cold clouds in transient spiral arm-like features, as observed e.g. in the Hubble Ultra Deep Field at $1<z<2$ \citep{ElmegreenElmegreen2014}. This indicates that the disk may have entered an epoch of ``disk settling'', as seen in the DEEP2 Survey \citep{Kassin2012} for galaxies of stellar mass $8.0< \log M_\star (M_\odot) < 10.7$ over $0.2 < z < 1.2$. We detect individual hot super bubbles from correlated feedback events in the extended disk, leading to galactic outflows of enriched gas. Most of this enriched gas is found to enter a galactic fountain, rather than large scale outflows as seen at higher redshifts ($z>2$). The gas disk metallicity is close to solar. 

The boosted feedback, necessary to regulate the rate of star formation in the {\tt ALL\_Efb\_e001\_5ESN} simulation, produces a significantly more turbulent system, with metals being spread over greater distances in comparison to the fiducial run. The cold gas metallicity in the central disk is also much lower, $Z_{\rm gas}\sim 0.1-0.2\,Z_\odot$. The strong feedback hinders the formation of a thin gas disk, indicating a striking difference in galaxy evolution compared to our fiducial model. A similar result was found by \cite{Agertz2011} and \cite{Roskar2013}, where increasing the strength of feedback led to reasonable global characteristics at the cost of destroying the galactic disk. In \S\ref{sect:feedback_thin} we discuss this in greater detail, and show that the fiducial model at $z=0$ features a thin stellar and gaseous disks.

\subsection{Properties of galactic winds}
\label{sect:winds}

The differences in galaxy evolution can further be characterized via properties of galactic winds. In Figure \ref{fig:wind} we show the wind mass loading, defined as $\eta=\dot{m}_{\rm wind}/{\rm SFR}$, and average wind velocity as a function of circular velocity of the galaxy, here simply defined as $\vel_{\rm circ}(r=20 \kpc)$, for {\tt ALL\_Efb\_e001}, {\tt ALL\_Efb\_e010} and {\tt ALL\_Efb\_e001\_5ESN}. Each point represents the main progenitor over cosmic time ($1\lesssim z\lesssim 7$). 

We compute the radial mass outflow rate in concentric shells via
\begin{equation}
\dot{m}=\sum_i^{N_{\rm cell}} m_i \vel_{i,{\rm rad}}/\Delta l,
\end{equation}
where $N_{\rm cell}$ is the number of cells in a shell, $m_i$ and  $\vel_{i,{\rm rad}}$ are the mass and radial velocity of the cell, and $\Delta l$ is the shell thickness, here typically in the range 100-200 pc. We only consider gas with radial velocities $\vel_{\rm rad}\ge 10\kmsec$, i.e. \emph{only} outflowing gas, and omit all gas belonging to the galactic disk by neglecting gas within a slab of thickness $\pm 2 \kpc$ encompassing the ISM in the disk plane. The outflow rate used to compute the mass loading is the average rate, i.e. $\dot{m}_{\rm wind}=\langle \dot{m}\rangle$ for all shells at radial distances $2\,\kpc\le r\le20\,\kpc$, and the characteristic outflow velocity ($\vel_{\rm wind}$) is computed in an analogous fashion.

As seen in Figure \ref{fig:wind}, the different models give rise to markedly different mass loading factors. The weak effect of feedback in {\tt ALL\_Efb\_e001} results in $\eta< 1$ at all times, leading to the significant overproduction of stellar mass discussed in \S\,\ref{sect:smhm}. The fiducial model ({\tt ALL\_Efb\_e010}) shows large mass loading factors of the order of $\eta\sim 10$ at low $\vel_{\rm circ}$, and hence at early times of galactic evolution. The mass loading here decreases with increasing galactic mass, as predicted by models based on momentum or energy driven winds; $\dot{m}_{\rm wind}/{\rm SFR}\propto \vel_{\rm c}^{-\alpha}$, where $\alpha\sim1-2$, \cite[e.g.][]{oppenheimerdave06}, although our simulated galaxy exhibits a more complex behaviour, not described by a single power law. We emphasize that the wind properties shown in these figures have resulted from hydrodynamics of gas flows between the scale of energy and momentum injection, $\sim 50-100$ pc, and the scale of measurement, $\sim 2-20$  kpc, and are therefore predictions of the simulations and not a result of any galactic wind model assumptions \citep[see e.g.][]{Okamoto2010,Vogelsberger2013}. 

Similar to our fiducial model, the {\tt ALL\_Efb\_e001\_5ESN} run produces large mass loading factors at low galactic masses, leading to similar SFRs compared to the fiducial model for $z\gtrsim 2$ (see \S\,\ref{sect:SFH}). However, values of $\eta\sim 10$ are present throughout the galactic evolution, even for $\vel_{\rm c}\gtrsim 200\,\kmsec$, hindering the formation of a thin galactic disk.

In the right hand panel of Figure \ref{fig:wind} we show the average radial wind velocities as a function of galaxy circular velocity for the three simulations, together with data derived from Na I D absorption measurements from \cite{SchwartzMartin2004} and \cite{Rupke2005b}, as well as from Mg\,II absorption lines from a sample of galaxies at $0.3<z<1.4$ \citep{Rubin2014}. Although the scatter is large, owing to the bursty nature of star formation and feedback, especially at low galaxy masses, it is intriguing that all simulations produce roughly the same wind velocities, close to the circular velocity of the galaxy, in broad agreement with observed outflow velocities. We leave a more detailed investigation of galactic wind properties for future work.

\subsection{The $\Sigma_{\rm gas}-\Sigma_{\rm SFR}$ relation}
\label{sect:KS}

In Figure \ref{fig:KS1} we plot the $\Sigma_{\rm gas}-\Sigma_{\rm SFR}$-relation (the Kennicutt-Schmidt (KS) relation) at $z=2-3$ for the entire simulation suite. We consider only the surface density of cold ($T\le10^4\K$) atomic and molecular gas, and do not include any contribution from helium\footnote{to scale our quoted surface densities to account for helium, multiply them by a factor of 1.36.}. We calculate $\Sigma_{\rm gas}$ and $\Sigma_{\rm SFR}$ in patches with an area $A=750\times 750\pc^2$ evenly distributed across the simulated disks. We define $\Sigma_{\rm SFR}\equiv m_\star t_\star^{-1}A^{-1}$, and consider the total mass of young stars $m_\star$ formed within $t_\star=20\Myr$. For each galaxy and redshift we bin the resulting relation for $\Sigma_{\rm gas}$ in logarithmic bin sizes of 0.15 dex, and each panel contains measurements of simulations snapshots in the redshift range $z=2-3$ at expansion factor intervals $\Delta a=0.01$. 

We compare the simulated relation to the KS relation inferred from observations of $z\sim 1-3$ normal star-forming galaxies from \cite{Genzel2010}, $z\sim 3$ Damped Ly-$\alpha$ Systems \citep[DLAs,][]{WolfeChen06}, $z\sim 3$ low surface brightness emission around Lyman break galaxies \citep[LBGs,][]{Rafelski2011} as well as the relation of \cite{kennicutt98} for $z\approx 0$ galaxies.  Our simulated galaxy is hosted by a halo of mass $M_{\rm h}\sim$ few $\times 10^{11}\Msol$ at $z\sim 3$, consistent recent constraints from the cross-correlation between DLAs and the Ly$\alpha$ forest that indicate that most DLAs at $z\approx 2-3$ are hosted by relatively massive halos \citep{Font2012}. At $2<z<4$, DLAs are observed to have a wide distribution of metallicities, $\log(Z_{\rm gas})\sim$ $-2.5$ to $-0.5$, with a peak around $\log(Z_{\rm gas})\sim$ $-1.5$ \citep{Prochaska2007}. Numerical models by \cite{Pontzen2008} have indicated that the metal rich DLAs are likely to be associated with halos of mass $M_{\rm h}\gtrsim 10^{10}\,\Msol$. 

The KS relation for the MW progenitor galaxy in the fiducial simulation ({\tt ALL\_Efb\_e010}) shown in the top left panel of Figure \ref{fig:KS1} is in agreement with the empirical KS relation for $\Sigma_{\rm gas}\gtrsim 100\Msol{\rm pc}^{-2}$, and shows a clear drop below this surface density. This transition surface density is related to the physical density at which molecular hydrogen can be synthesized on dust grains \citep{schaye01,Gnedin09,kmt09,GnedinKravtsov2010,GnedinKravtsov11}. The lower star formation efficiency below this transition surface density, where the simulations match the DLA and LBG data, arises from the low gas metallicity, $Z_{\rm g}\sim0.1-0.2\,Z_{\odot}$, in the outer disk which in turn results in a low $f_{\HH}$. 

The KS relation in the simulation with $\epsilon_{\rm ff}=10\%$ ({\tt ALL\_Efb\_e010}) is hence in very good agreement with the observed KS relation of both low-$z$ and high-$z$ galaxies. Note that simulations with identical ingredients, but with $\epsilon_{\rm ff}=1\%$ ({\tt ALL\_Efb\_e001}), is also consistent with observations at high surface densities but exhibits a drop in star formation at a somewhat larger gas surface density. The fact that the normalization of the KS relation is similar in simulations with a {\it local} efficiency of star formation different by a factor of ten illustrates that in simulations with efficient feedback, and significant outflows, the normalization of the KS relation does not reflect the local star formation efficiency. In this case, the global star formation rate self-regulates to produce a low overall star formation efficiency (i.e., long gas consumption time scales). 

In contrast, in simulations in which feedback is weak or absent, the normalization of the KS relation is linearly related to the local efficiency. Thus, for example, the normalization in the simulation with $\epsilon_{\rm ff}=10\%$, but feedback turned off, is approximately an order of magnitude larger than that in the simulation with  $\epsilon_{\rm ff}=1\%$ for $\Sigma_{\rm gas}\gtrsim 100\Msol{\rm pc}^{-2}$.

Figure \ref{fig:KS1} shows that the simulation with $\epsilon_{\rm ff}=1\%$ and the SN energy output boosted by a factor of five ({\tt ALL\_Efb\_e001\_5ESN}) has significantly lower normalization of the KS relation compared to our fiducial simulation and in tension with observations. As demonstrated more quantitatively in \S\,\ref{sect:massmet}, this arises due to the very efficient removal of metal rich gas from the galaxy, leaving the entire disk metal poor with an outer disk metallicity $Z_{\rm g}<0.1Z_{\odot}$. The marked difference between the internal star formation properties of {\tt ALL\_Efb\_e010} and {\tt ALL\_Efb\_e001\_5ESN}, which both conform to all observed \emph{global} galactic characteristics, illustrates that it is potentially possible to break the degeneracy between such models using additional properties and observations. 

\section{Discussion}
\label{sect:discussion}

\subsection{Comparison with previous studies}

A wide range of numerical studies of galaxy formation focusing on different stellar feedback processes have appeared in the past several years. It is thus useful to discuss how these models differ from, or agree with, the numerical models presented in this work and why.

Recently, \cite{Hopkins2014} presented a series of high-resolution cosmological zoom-in smoothed particle hydrodynamics (SPH) simulations of galaxy formation run to $z = 0$, spanning halo masses $M_{\rm halo}\sim 10^8-10^{13}\,M_\odot$. Our results generally agree with those of Hopkins et al., who also find that the star formation efficiency tends to self-regulate in the regime when stellar feedback is efficient. In particular, they found that the observed low normalization of the KS relation was reproduced in their simulations even when a local star formation efficiency as high as $\epsilon_{\rm ff}=100\%$ was used. 
Furthermore, \cite{Hopkins2014} also conclude that both early radiative feedback and subsequent supernova feedback are important. For the latter they use a scheme that captures the momentum generated during the  (often unresolved) Sedov-Taylor stage of evolution. 

\cite{TrujilloGomez2013}, and previously \cite{Ceverino2013}, presented  adaptive mesh refinement (AMR) simulations of galaxy formation at high $z$, in the regime of dwarf galaxies ($M_{\rm halo}(z=0)=3\times10^{10}\,\Msol$) and low mass spiral galaxies ($M_{\rm halo}(z=0)=2\times 10^{11}\,\Msol$). Using an implementation similar to what is presented in this work, the authors demonstrated the importance of considering radiation pressure in galaxy formation simulations. However, at their current resolution ($40-80\,{\rm h}^{-1}$ pc at $z = 0$) the effect of thermal feedback is possibly underestimated, as indicated by the star formation histories in figure 8 in \cite{TrujilloGomez2013}, where the spiral galaxy's SFR is overpredicting the rates of \cite{Behroozi2013} by almost a dex at $z=1.5$. However, the authors compare their simulated galaxy to the data of a galaxy with half the dark matter halo mass. Accounting for this offset brings their model with strong radiative feedback into closer agreement with the semi-empirical expectations (within the $1 \sigma$ confidence interval: Trujillo-Gomez, private communication)

Using SPH simulations of galaxy formation in halos of masses in the range $M_{\rm halo}=10^{11}-3\times 10^{12}\,M_\odot$, \cite{Aumer2013} studied the impact of their feedback model based on the multiphase SPH code presented in \cite{Scannapieco06}, with the addition of  momentum input from radiation pressure. A good match to global galaxy characteristics at $z=0-4$, specifically for Milky Way analogues, was recovered if the authors considered a large value of the infrared optical depth, $\tau_{\rm IR}=25$, but allowed for more gentle momentum input in low redshift systems. The authors identified that despite the effort in tuning feedback parameters, the model still overpredicted the mass of stars formed at $z > 4$, and argued that this may be due to inaccurate modeling of star formation at early stages of galaxy formation, or simply due to the specific merger histories of the simulated haloes.

\cite{Brook2012} and \cite{Stinson2013} discussed the importance of ``early feedback'' in their SPH galaxy formation simulations. These authors assume that $10\%$ of the \emph{bolometric} luminosity radiated by young stars get converted into thermal energy, which significantly affected properties of their simulated galaxies. Although this model differs significantly from our subgrid model of radiation pressure, in which we consider the actual momentum transfer from radiation via local gas/dust UV and IR absorption, the concept of pre-supernovae feedback was shown to have a significant effect on galaxy evolution, in agreement with our conclusions.

All of the above authors have recognized the importance of additional feedback processes in addition to supernovae energy input, in particular  momentum injection due to radiative feedback that pre-conditions star-forming regions before the first supernovae explosion occurs ($t\sim 4\Myr$). This is indeed also the case in our models, where the lack of early momentum based feedback results in star formation rates that are a factor of 2-10 times higher than the average expected values, see Figure \ref{fig:SFH}.

\begin{figure*}
\begin{center}
\begin{tabular}{ccc}
\includegraphics[width=0.4\textwidth]{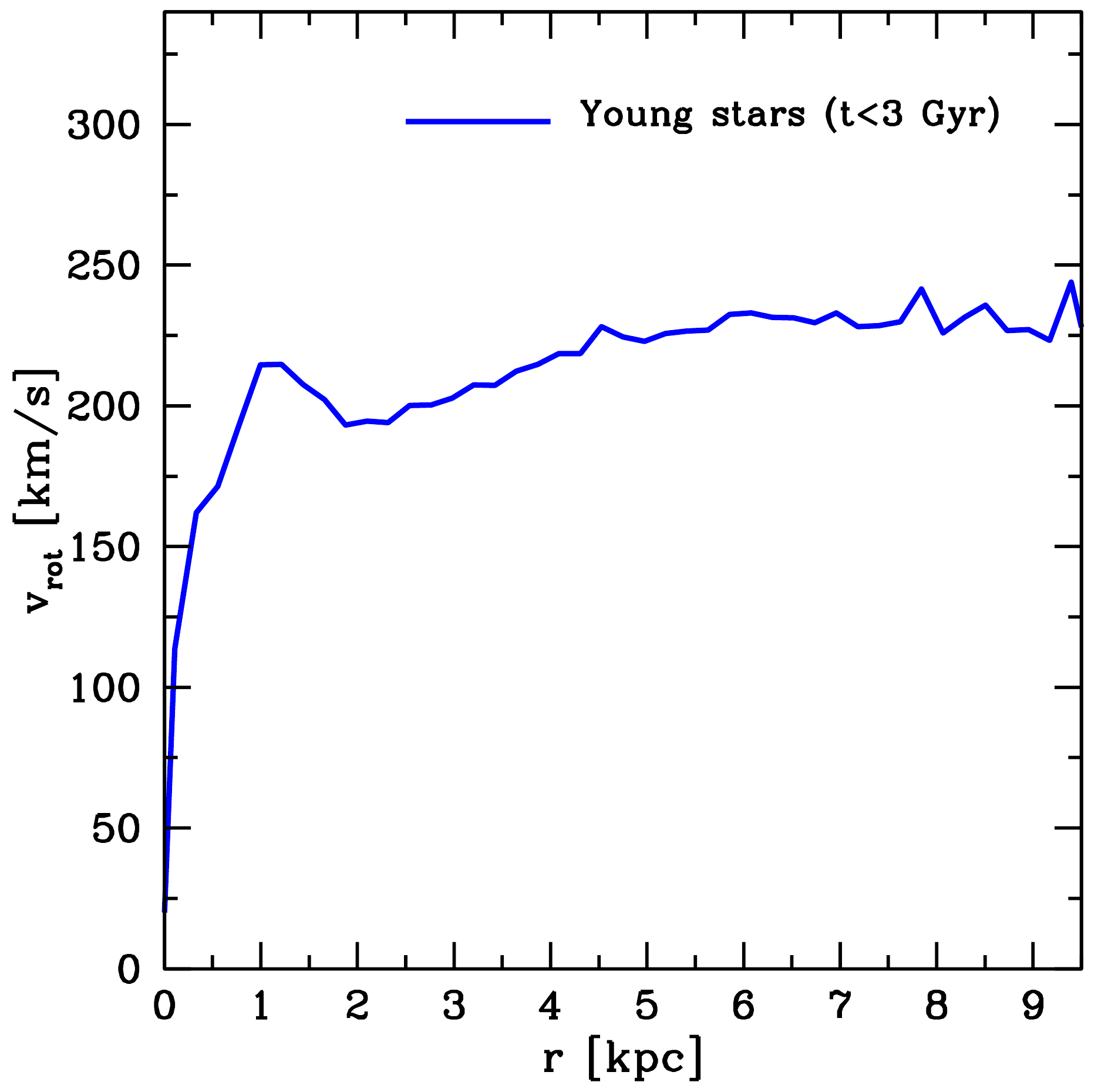}
\includegraphics[width=0.4\textwidth]{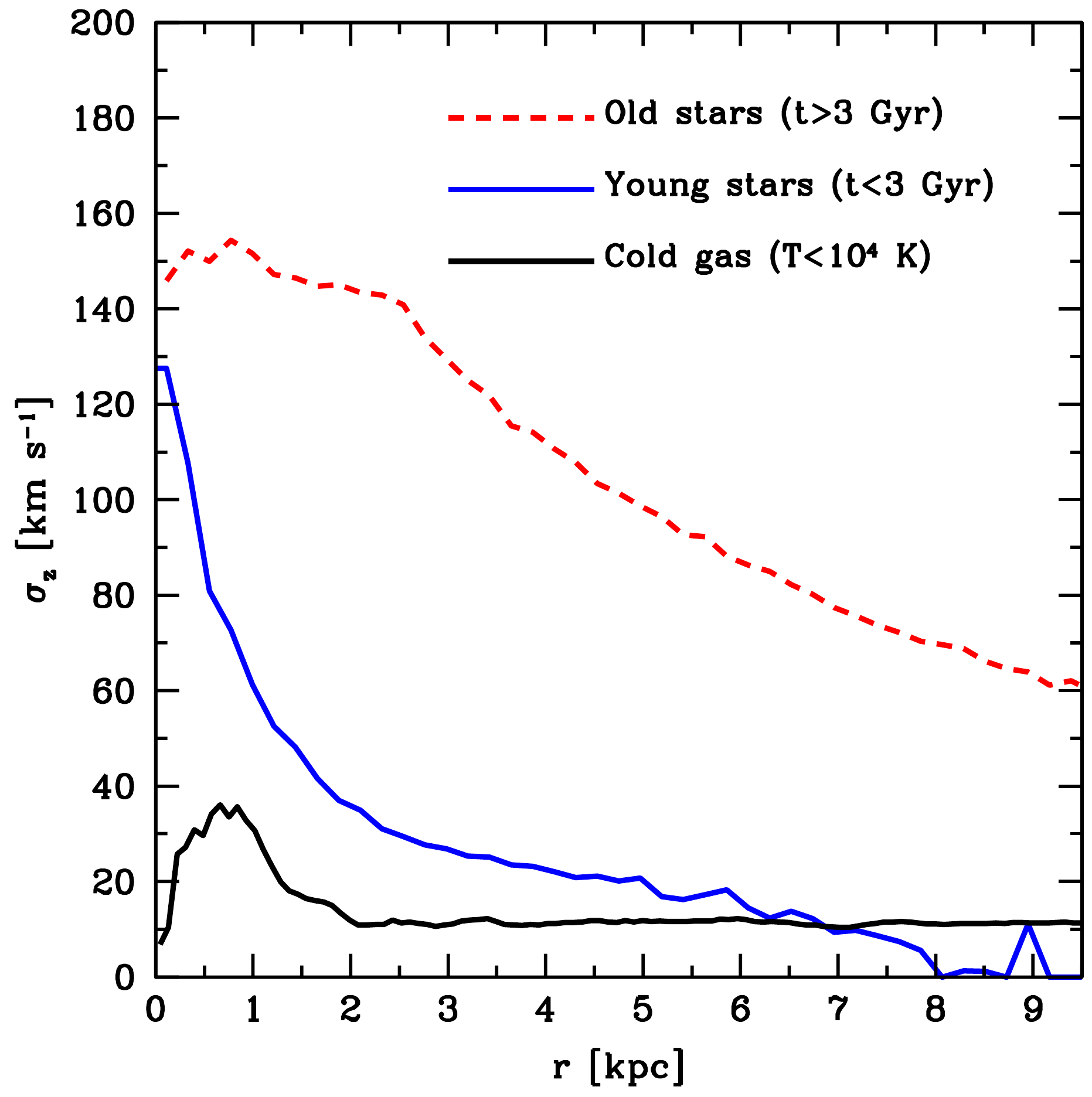}
\end{tabular}
\caption{(Left) The rotational velocity profile for young stars ($t<3$ Gyr) in the fiducial model {\tt ALL\_Efb\_e010} at $z=0$. (Right) Vertical velocity dispersion profiles for the young ($t<3\Gyr$) and old ($t>3\Gyr$) stars in the disk, as well as the cold ($T<10^4$ K) gas disk in the fiducial model {\tt ALL\_Efb\_e010} at $z=0$. Young stars reside in a kinematically colder configuration in comparison to the old stars. The specific values of $\vel_{\rm rot}/\sigma_z$ are in good agreement with those observed for the Milky Way, where young (low $[\alpha/{\rm Fe}]$) and old stars (high $[\alpha/{\rm Fe}]$) near the solar circle (at $r\sim 8\kpc$) have $\vel_{\rm rot}/\sigma_z\gtrsim 10$ and $\sim 4$ respectively \citep[see][]{Bovy2012}. This illustrates that our fiducial model does not suffer, at least to the same extent, of the issues raised by \cite{Roskar2013}, where efficient stellar feedback led to a complete destruction of any thin/cold galactic components.
}
\label{fig:z0disk}
\end{center}
\end{figure*}

Recently, \cite{Marinacci2014} presented cosmological simulations, using the {\tt AREPO} code \citep{arepo}, of eight Milky Way-sized haloes, previously studied using dark matter only in the Aquarius project \citep{Springel2008}. The simulated galaxies had realistic sizes, rotation curves and stellar-mass to halo-mass ratios, and the authors noted this was achieved without resorting to factors thought to be crucial for galaxy formation by earlier studies, e.g. a high density threshold for star formation \citep[e.g.][]{Governato10}, a low star formation efficiency \citep[][]{Agertz2011}, or early stellar feedback \citep[e.g.][]{Brook2012,Agertz2013,Hopkins2014}. While the models of \cite{Marinacci2014} demonstrate convincing resolution convergence, as well as encouraging galaxy properties, this neither negates previous work nor comes as a surprise; Marinacci et al. adopt a stellar feedback approach based on a kinetic wind scheme in which the wind velocity, and mass loading, is scaled with the local dark matter halo mass \citep{PuchweinSpringel2013}. At the adopted resolution ($\sim 340-680$ pc force softening), a direct modeling of strong feedback tends to affect too much gas due to mixing at the resolution scale, as well as the inability to resolve the multiphase ISM \citep{Roskar2013}. The type of wind scheme adopted by Marinacci et al. circumvents these problems by essentially postulating the existence of outflows. A natural benefit of this approach is a better resolution convergence on global galactic properties.

\subsection{$H_2$-based star formation and the efficiency of feedback}

In this work we have adopted a star formation model based on the local abundance of molecular hydrogen using the formalism presented by \cite{kmt09}. Previous work \citep{Gnedin09,GnedinKravtsov2010,GnedinKravtsov11,Kuhlen2012} have demonstrated how this approach leads to suppressed star formation in metal poor environments typical for dwarf galaxies, even resulting in a population of completely dark galaxies situated in dark matter haloes of mass $M_{\rm halo}\lesssim 10^{10}\Msol$ \citep{GnedinKravtsov2010,Kuhlen2013}. A relatively unexplored outcome of such H$_2$-based star formation model is its ability to boost feedback \citep[but see][]{Christensen2014}. In fact, most previous studies completely neglect feedback or include only inefficient thermal feedback from supernovae. 

Stellar feedback can be boosted in the H$_2$-based star formation model in several ways. For example, star formation can become more localized because in low-metallicity environments, high gas densities is required for vigorous star formation. This can lead to more correlated energy and momentum injection events. Indeed, this effect was pointed out by \cite{Christensen2014}. Furthermore, as the molecular hydrogen fraction, $f_{\rm H_2}$, is a function of the local dust abundance (and hence gas metallicity), rapid \emph{local} enrichment of gas from newly formed stars allows for a sudden decrease in the effective star formation threshold, leading to local burst of star formation with associated strongly correlated feedback. As we have demonstrate in this study (see Appendix\,\ref{appendix:A}), this changes the nature of the star formation-feedback cycle, resulting in a more efficient suppression of star formation  compared to models with fixed star formation density thresholds (here $n=25\cc$). We note that once the numerical resolution is sufficiently high to allow for star formation to robustly occur at densities $n\gg100\cc$, at which gas is expected to be mostly molecular, an explicit subgrid model for $f_{\rm H_2}$ may have little effect over a fixed high density threshold, as argued by \cite{Hopkins2012structure}.

\subsection{Feedback and the survivability of a thin galactic component}
\label{sect:feedback_thin}

As discussed in the above sections, efficient feedback leading to galactic outflows is a necessary ingredient in order to match a large number of global observables. A caveat to this was raised by \cite{Roskar2013} who demonstrated that while strong feedback can produce stellar masses that conform to semi-empirical $M_\star-M_{\rm halo}$ relations from e.g. \cite{Behroozi2013} and \cite{Moster2013}, this had severe consequences on the final galactic disk; no thin stellar disk nor cold gaseous disk survived. In Figure \ref{fig:z0disk} we show the $z=0$ rotational velocity and velocity dispersions of young and old stars as well as cold gas. Based on the mono-abundance population data by \cite{Bovy2012}, Ro{\v s}kar et al. raised the point that $\vel_{\rm rot}/\sigma$, at least for the Milky Way, is $\gtrsim 10$ for young stars (taken to be stars younger than $3\Gyr$) at the solar radius, and closer to $\sim 4$ for older stars. Using the same age cut we find that we do not suffer, at least to the same extent, from the problem of Ro{\v s}kar et al; we clearly see a young thin stellar component with a velocity dispersion at the solar radius close of $\sim 0-20\kmsec$. A cold gaseous disk is present with velocity dispersions of $\sim 10\kmsec$ outside of the bulge ($r\gtrsim 1.5\kpc$), typical of local spiral galaxies \citep{Tamburro2009,Agertz09}. We note that this does not mean we are not suffering from numerical heating due to low resolution, or that the thin disk is the dominant galactic component, only that our approach to feedback and star formation does not necessarily lead to disk destruction as found in \cite{Roskar2013}.

\subsection{Caveats, small scale issues, and the next step}
\label{sect:caveats}

\paragraph{The efficiency of star formation} Although our simulation suite was carried out with relatively high numerical resolution, most key processes related to feedback and star formation remain subgrid, as they operate on $\sim$ pc scales within GMCs. Given that the true density probability distribution function (PDF) relevant for star formation is not fully resolved in galaxy formation simulations, the adopted star formation efficiencies per free fall time may be modified at higher resolution.  The local gas depletion time is assumed to be $t_{\rm SF}=t_{\rm ff}/\epsilon_{\rm ff}$, and is only modeled, and measured, on large scales ($\gtrsim 100$ pc), and the adopted value of $\epsilon_{\rm ff}$ discussed in this work hence only applies on these scales \citep[see][for a recent discussion of how $t_{\rm SF}$ may manifest on different scales]{Gnedin2014}.

A number analytical and numerical studies of star formation in super sonic turbulence, aimed at understanding what sets the star formation efficiency per free-fall time and its evolution in GMCs, has been carried out recently \citep[see e.g.][]{KrumholzMcKee2005,PadoanNordlund2011,Semadeni2011,Hennebelle2011}. These studies find that the value of $\epsilon_{\rm ff}$ depends on detailed properties of star forming clouds, e.g. the flow Mach number as well as the virial parameter\footnote{the ratio of the cloud kinetic energy to gravitational potential energy}, $\alpha_{\rm vir}$ \citep[see review by][]{Padoanreview2013}, all leading to a time dependent density PDF where stars form in the high density tail consisting of molecular clumps ($n\sim10^2-10^4\cc$) and cores ($n\gtrsim 10^5\cc$). A generic result is that unless star formation is regulated by radiative feedback, protostellar outflows, subsequent supernovae as well as magnetic fields, the resulting efficiency can be significantly larger than the $\epsilon_{\rm ff}\sim 0.25-0.5\%$ deduced from observations on kpc scales (see \S~\ref{sec:sfeff}), especially for gravitationally bound clouds ($\alpha_{\rm vir}<1$).

To some degree, the assumption made in our work regarding feedback regulated star formation is in line with the above results, although we apply the efficiency on much larger, $\sim 100$ pc, scales. More work is definitely necessary in order to ``connect the scales'', and future improvements in numerical resolution should allow  \emph{global} characteristic of star forming regions, such as the virial parameters $\alpha_{\rm vir}$, to be at least marginally resolved. \cite{Padoan2012} demonstrated \citep[but see][]{GonzalezSamaniego2013} how the measured star formation efficiency per free fall time in high resolution simulations of supersonic turbulence could be expressed as a simple law depending only on the cloud free-fall and dynamical time, $\epsilon_{\rm ff}\propto \exp(-1.6\,t_{\rm ff}/t_{\rm dyn})$. It remains to be seen whether this kind of assumption propagates to differences in large scale galactic observables in comparison to the choice of a large uniform $\epsilon_{\rm ff}$ ($\sim 10\%$) in our work \citep[see also][]{Hopkins2014}. 

\paragraph{The star formation recipe} The assumption of an underlying non-linear star formation law (here $\dot{\rho_\star}\propto \rho_{\rm gas}^{1.5}$) may be incorrect. \cite{Gnedin2014} argued that such small scale relation should result in a non-linear slope in the observed KS relation, which is incompatible with the linear relation observed for molecular gas ($\Sigma_{\rm SFR}\propto\Sigma_{\rm H_2}$) at $\Sigma_{\rm gas}\gtrsim 100\, M_{\odot}\,\pc^{-2}$ in the THINGS survey \citep{bigiel2008}. Although our fiducial model is in good agreement with the observed KS relation at $z\sim 1-3$, it remains to be seen if the same feedback models can regulate star formation at high surface densities to be compatible with local observations of starbursts.

\paragraph{Modeling thermal feedback} Our stellar feedback model accounts for radiation pressure, stellar winds and supernovae type II and Ia, as well as associated mass loss and metal generation where appropriate \citep{Agertz2013}.  While the overall energy and momentum budget has been shown to generate realistic galaxy properties, at least for $z\gtrsim1$, the detailed role of a separate feedback energy variable remains to be explored. It is clear that storing even a small fraction of SN energy in such variable significantly affects the star formation rate and efficacy of feedback. It is thus important to understand in detail the physical nature of such extra energy component. 

As mentioned in \S\,\ref{sect:FB}, this variable can be viewed as accounting for the effective pressure from a multiphase medium, where local unresolved pockets of hot gas exert work on the surrounding cold phase. Alternatively,  it can interpreted as crudely modeling kinetic energy stored in unresolved small-scale turbulence or in cosmic rays (CRs). Indeed, \cite{Booth2013} \citep[see also][]{Hanasz2013,SalemBryan2014} demonstrated that if a modest fraction of the available supernova energy ($\sim 10\%$ of $10^{51}$ erg) is injected as a CR energy density, galactic winds can be driven effectively and can exhibit qualitatively different properties compared to SN driven winds. In future work we will explore models in which cosmic ray feedback contribution to the stellar feedback budget is modeled explicitly in a fully cosmological setting.

\paragraph{Numerical resolution and convergence} We note that the star formation and feedback recipes used in our simulations have been specifically designed to operate on scales of $\sim 50-100$ pc or below, comparable to the sizes of massive GMCs, as we discuss in detail in our previous paper \citep{Agertz2013}. We therefore expect them to work best at this particular resolution. In fact, running our fiducial model at $\Delta x\sim300$ pc resolution produces a different star formation history (not presented here), with more stars formed at early times compared to the models presented in this paper. This is not surprising, as at lower resolution, stellar feedback is spread over much more mass, hence achieving lower heating/momentum injection rates, leading to weaker galactic winds. 

Of course, it would be desirable for cosmological simulations to be as insensitive to the numerical resolution as possible. However, because the density PDF of the ISM changes with resolution, it is not guaranteed that a specific model of the star formation/feedback cycle is invariant to the gas density PDF change. Simulations where such models are tied to converged quantities, e.g. properties of galactic winds depend the total mass of the host dark matter halo as in \cite{Marinacci2014} \citep[see also][]{Vogelsberger2014}, are naturally less sensitive to changes in the numerical resolution, as discussed above.

\section{Conclusions}
\label{sect:conclusions}
In this paper we have presented a suite of high resolution cosmological zoom-in simulations of galaxy formation, focusing on the formation of a Milky Way-sized galaxy with a halo mass of $M_{\rm 200}\approx 10^{12}\Msol$ at $z=0$. We have focused on exploring how variations in the modeled star formation and feedback physics affect galaxy evolution and how properties of the simulated MW progenitors compare to modern high redshift ($z\gtrsim1$) estimates of global characteristics, such as star formation histories, the mass-metallicity relation, the Kennicutt-Schmidt relation and the stellar mass - halo mass relation. Our simulations adopt the feedback model presented recently by \cite{Agertz2013}, which accounts for energy and momentum injection via radiation pressure, stellar winds and supernovae type II and Ia. Furthermore, star formation is modeled using the local density of molecular gas \citep{kmt09,Gnedin09,Kuhlen2012}. 

Perhaps the central result of our study is that in our implementation feedback becomes efficient in suppressing star formation and driving outflows only if the {\it local} star formation efficiency per free fall time is sufficiently large, $\epsilon_{\rm ff}\approx 10\%$ for the density field resolved in the current simulations. Such large efficiency allows for a high degree of temporal and spatial correlation of energy and momentum injection. We show (in \S~\ref{sec:sfeff}) that such value of the local efficiency is consistent with observational estimates in giant molecular clouds. We confirm that in the models with efficient feedback, the star formation efficiency measured on global, kiloparsec scales self-regulates to the low value inferred from observations. 

The rest of our results can be summarized as follows.

\begin{itemize}
\item At the peak spatial resolution of our simulations, $\Delta x\sim75$ pc, simulated galaxy relations are sensitive not only to the details of stellar feedback processes and their parameters, but also to the underlying star formation model and the adopted efficiency of star formation. This highlights the fact that it is important to model carefully the entire star formation and feedback cycle. 

\item If the adopted $\epsilon_{\rm ff}$ is low (here $\sim 1\%$, relevant for the currently resolved gas density field), hence treating the observed inefficiency of galactic star formation as a model input rather than a prediction from the star formation--feedback cycle, the strength of feedback must be artificially boosted in order to regulate galaxy masses via galactic outflows. We show that although this can lead to a successful match to the semi-empirical stellar mass--halo mass relation, such simulations may be in tension with the normalization of the Kennicutt-Schmidt relation. Furthermore, in agreement with other recent studies \citep{Agertz2011,Roskar2013}, we also find that simply boosting feedback with a low $\epsilon_{\rm ff}$, to match global relations, prevents the formation of a well-defined gaseous disk, even at relatively low redshifts  ($z\lesssim 1$). The morphology of the gaseous and stellar galactic disks may therefore serve as one of the key additional constraints on the parameters of the star formation--feedback loop. 

\item Our simulations indicate a complex interplay between the parameters of star formation and stellar feedback. If the star formation efficiency is sufficiently large to allow for feedback self-regulation, removing key feedback sources, such as radiation pressure or efficient thermal feedback, moves the galaxy off observed scaling relations, but in a complex manner. 

\item Encouragingly, we find that our fiducial model provides a good match to all considered observables at different redshifts: semi-empirically derived star formation histories, the stellar mass-gas metallicity relation and its evolution, the Kennicutt-Schmidt relation, the $M_\star-M_{\rm halo}$ relation and its evolution, as well as the flat shape of rotation curves and galaxy morphology. 
In particular, we show that our fiducial simulation, with feedback sufficient to drive vigorous galactic winds at high-$z$, is sufficiently gentle to allow for a young thin stellar disk to form by $z=0$. The disk has a flat rotation curve, with gas and stellar velocity dispersions consistent with observations of the Milky Way's at the solar circle.
\end{itemize}

Our results are encouraging, as they show that a comprehensive model that satisfies a number of non-trivial observational constraints and tests is feasible. In this work we have mostly discussed the $z\gtrsim 1$ results for our simulations, as the majority of them were stopped at high redshifts due to the computational expense. A significant fraction of stars in the $z=0$ thin disk is expected to form after $z\sim 1$ \citep{vandokkum2013}, which also appears to be the case in our fiducial model where we see the formation of well-defined thin stellar disk as soon as the turbulent gas rich disk enters an epoch of ``disk settling'' \citep{Kassin2012} at $z\lesssim 1$ (see Figure \ref{fig:map2}, \ref{fig:z0disk} and related text). Building upon the exploratory study presented here, we will in future work (Agertz \& Kravtsov, in prep) study and contrast galaxy sizes and morphologies at $z=0$ for a subset of the simulated galaxies.

\acknowledgements
The simulations presented in this paper have been carried using the Midway cluster at the University of Chicago Research Computing Center.
We would like to thank Douglas Rudd for his support in running the simulations. We thank Romain Teyssier, Phil Hopkins and Du{\v s}an Kere{\v s} for fruitful discussions. AK would like to thank the Simons foundation and organizers and participants of the Simons symposium on Galactic Super Winds in March, 2014, for stimulating and helpful discussions that aided in preparation of this paper. AK  was supported via NSF grant OCI-0904482, by NASA ATP grant NNH12ZDA001N, and  by the Kavli Institute for Cosmological Physics at the University of Chicago through grants NSF PHY-0551142 and PHY-1125897 and an endowment from the Kavli Foundation and its founder Fred Kavli.

\bibliographystyle{apj}
\bibliography{cosmo.bbl}

\appendix
\section{Impact of a fixed density threshold for star formation}
\label{appendix:A}
\begin{figure}[ht]
\begin{center}
\includegraphics[scale=0.5]{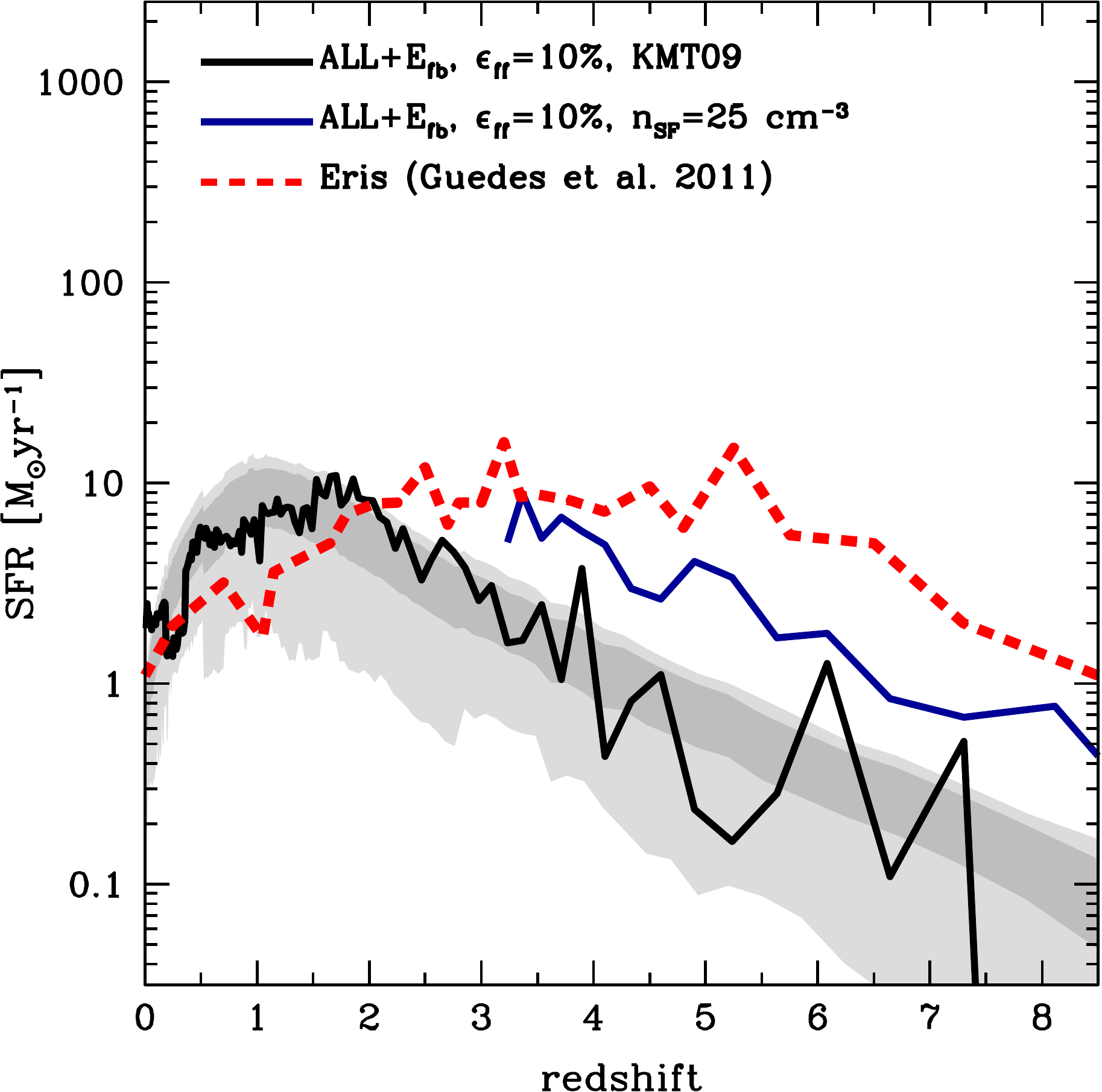}
\caption{Simulated star formation histories compared to the \cite{Behroozi2013} data for $M_{\rm vir}(z=0)=10^{12}\Msol$. Dark and light gray shaded areas are one-and two-sigma confidence regions respectively. We adopt bins of size $\Delta t_{\rm SF}=100\Myr$ for the simulated SFHs. The KMT09 model shows SFRs lower by$\sim 0.5$ dex  compared to the fixed density threshold model. For comparison we also plot the SFH of the Eris simulation \citep[][]{Guedes2011}.}
\label{fig:SFHn25}
\end{center}
\end{figure}

The assumed star formation model throughout the is work is based on the abundance of molecular hydrogen, see Equation\,\ref{eq:schmidtH2}. A more common approach in the galaxy formation community is to only allow stars to form above some fixed density threshold $\rho_{\star}$, i.e.
\begin{equation}
\label{eq:schmidt}
\dot{\rho}_{\star}=\frac{\rho_{\rm g}}{t_{\rm SF}}\,\,{\rm for} \,\,\rho>\rho_{\star}.
\end{equation}
The appropriate value of this threshold is highly resolution dependent, as galaxy formation simulations do not yet converge on a density PDF representative of the ISM,  and values of this threshold greatly vary in the literature \citep[see discussion in][]{Agertz2011}. To study the impact of the star formation prescription choice, we adopt the density threshold of $\rho_\star=25\cc$, which roughly corresponds to the physical density at which $f_{\HH}\sim 50\%$ at $Z_{\rm g}=Z_\odot$ \citep{Gnedin09}. Figure \ref{fig:SFHn25} shows dependence of the star formation history of the main progenitor on changes of the star formation prescription {\it only}, from the molecular based prescription adopted in our study to the fixed density threshold prescription of Equation\,\ref{eq:schmidt}. All the other feedback and star formation related settings, e.g. the efficiency per-free-fall were fixed at their fiducial values. The H$_2$ model results in SFRs lower by $\sim 0.5$ dex for $z<7$ compared to the traditional constant density threshold model. The latter disagrees with the Behroozi et al inference at all redshifts. As shown in the figure, the constant threshold model is in fairly good agreement with the SFH of the Eris simulation \citep[][]{Guedes2011}, which also adopted the constant threshold based star formation model, but with the threshold of $n_{\star}=5\cc$. Note that the virial mass of the Eris simulation ($M_{\rm vir}\approx 7.9\times 10^{11}\,\Msol$) is $\sim 20\%$ lower than the simulated halo in this study. 

\end{document}